\documentclass[a4paper]{article}  
\usepackage[margin=1in]{geometry}

\usepackage{amsmath}
\usepackage{amsfonts}
\usepackage{amsthm}
\usepackage{amssymb}
\usepackage{mathrsfs}

\usepackage{graphicx}
\usepackage{subfigure}
\usepackage{comment}

\usepackage{quantikz}
\usepackage{tikz}
\usetikzlibrary{arrows.meta,calc}

\usepackage{algorithm}
\usepackage{algpseudocode}
\usepackage{listings}
\usepackage{xstring}

\usepackage{hyperref}
\hypersetup{
    colorlinks = true,
    bookmarksopen = true,
    bookmarksnumbered = true,
    linkcolor = blue,
    anchorcolor = blue,
    citecolor = blue,
}
\usepackage[capitalise, nameinlink]{cleveref}

\newtheorem{definition}{Definition}

\newtheorem{lemma}{Lemma}
\newtheorem{corollary}{Corollary}
\newtheorem{theorem}{Theorem}
\newtheorem{remark}{Remark}

\newcommand{\veps}{\varepsilon}

\newcommand{\ri}{\mathrm{i}}
\newcommand{\re}{\mathrm{e}}
\newcommand{\rd}{\mathrm{d}}

\newcommand{\frakP}{\mathfrak{P}}
\newcommand{\scrH}{\mathscr{H}}
\newcommand{\sfJ}{\mathsf{J}}

\newcommand{\calT}{\mathcal{T}}
\newcommand{\calE}{\mathcal{E}}
\newcommand{\calL}{\mathcal{L}}
\newcommand{\calU}{\mathcal{U}}

\newcommand{\calR}{\mathcal{R}}

\newcommand{\calV}{\mathcal{V}}
\newcommand{\calW}{\mathcal{W}}
\newcommand{\calA}{\mathcal{A}}

\newcommand{\hlambda}{\hat{\lambda}}

\newcommand{\bzero}{\boldsymbol{0}}

\newcommand{\tprime}{t^{\prime}}

\newcommand{\tr}{\mathrm{Tr}}
\newcommand{\Lonenorm}[1]{\left\|#1\right\|_{L^1}}
\newcommand{\tracenorm}[1]{\left\|#1\right\|_{\mathrm{tr}}}
\newcommand{\superopnorm}[1]{\left\|#1\right\|_{1\to 1}} 
\newcommand{\opnorm}[1]{\left\|#1\right\|}

\newcommand{\dil}{\mathrm{dil}}

\newcommand{\rank}{\mathrm{rank}}
\newcommand{\odd}{\mathrm{odd}}
\newcommand{\even}{\mathrm{even}}
\newcommand{\trot}{\mathrm{Trot}}

\begin{document}

\title{Quantum Simulation of Non-Markovian Open Quantum Systems via Low-Rank Kernel Compression and Coupling Dilution}

\date{}

\author{Hengzhun Chen\thanks{School of Mathematical Sciences, Fudan University
    (\href{mailto:hzchen23@m.fudan.edu.cn}{\texttt{hzchen23@m.fudan.edu.cn}}).}
\and Zhenning Cai\thanks{Department of Mathematics, National University of Singapore
    (\href{mailto:matcz@nus.edu.sg}{\texttt{matcz@nus.edu.sg}}).}
}

\maketitle

\begin{abstract}

We present an error-controlled framework for the quantum simulation of
non-Markovian open-system dynamics in Gaussian environments. For a
system coupled linearly to a Gaussian environment with zero mean, the
bath's influence on the reduced dynamics is completely determined by its
two-time correlation kernel. This allows us to replace the microscopic
environment with an equivalent bath that has the same kernel. On a
finite simulation interval, a rank-$N$ approximation of the kernel
yields an effective bath of $N$ bosonic modes with time-dependent
couplings to the system. To encode these modes in qubits, we first
truncate each mode to its lowest $d$ occupation levels. When stronger
couplings lead to large occupations, we introduce coupling dilution,
which replaces each mode with several more weakly coupled copies while
preserving the bath correlation kernel. This distributes excitations
across more modes and trades additional modes for smaller local
dimensions. We derive explicit reduced-state error bounds and establish
polynomial circuit width and depth in simulation time and inverse
accuracy under stated spectral-decay and coupling-regularity assumptions,
provided elementary unitaries admit polynomial-cost implementations.

\end{abstract}

\section{Introduction}

Open quantum systems arise when a quantum system we are interested in
has non-negligible interactions with its uninteresting environment. The
composite system evolves unitarily, while the system alone is described
by a reduced density operator obtained by tracing out the environment.
This reduced dynamics is generally non-unitary and captures physical
effects such as dissipation, relaxation, decoherence, and information
exchange with the bath \cite{breuer2002theory}. Open-system descriptions
are therefore central to a wide range of quantum problems, including
quantum computing \cite{pellizzari1995decoherence,temme2017error},
quantum information processing \cite{nielsen2010quantum}, quantum
control \cite{pechen2014incoherent,wiseman2009quantum}, and chemical
physics \cite{weiss2012quantum}. They also appear as effective
descriptions in many-body methods \cite{gull2011continuous}. The
accurate simulation of open quantum dynamics is consequently a basic
task across quantum science and technology.

The numerical simulation of open quantum dynamics is challenging because
the system of interest is coupled to a bath with many, and often
continuum, degrees of freedom.  Directly resolving the full system--bath
state is usually infeasible, while eliminating the bath leads to reduced
dynamics with dissipation, fluctuations, and possibly long-lived memory.
In weak-coupling or short-memory regimes, the environmental degrees of
freedom can be incorporated into a reduced master equation.  Examples
range from Redfield-type equations \cite{redfield1965theory} to the
Markovian Lindblad equation
\cite{gorini1976completely,lindblad1976generators}.  Such time-local
descriptions are computationally economical, but their underlying
approximations may fail when the coupling is strong or the bath
correlation time is comparable to the characteristic system time scale.
Non-Markovian methods instead retain environmental memory more
explicitly.  These include Nakajima--Zwanzig memory-kernel equations
\cite{nakajima1958quantum,zwanzig1960ensemble}, influence-functional and
path-integral methods \cite{feynman2000theory,makri1995tensor},
hierarchical equations of motion \cite{tanimura1989time}, and
tensor-network simulations of chain-mapped environments
\cite{chin2011chain,prior2010efficient}.  Although powerful, their
computational cost can grow rapidly with the system Hilbert-space
dimension, the bath memory time or the temperature structure. Accurately
treating non-Markovian dynamics therefore remains particularly demanding
\cite{devega2017dynamics}.

In this context, quantum computers provide an alternative approach for
simulating the open-system dynamics. Quantum algorithms for open-system
dynamics may be classified by how they manage nonunitarity and
environmental information. One class embeds the dynamics into a larger
unitary evolution, using dilation ancillas or an explicit environment
register \cite{ding2024simulating,wang2011quantum}. Another class starts
from an equation for the reduced system dynamics. For Lindblad
evolution, examples include product-formula algorithms for local
Liouvillians \cite{kliesch2011dissipative} and algorithms based on
linear combinations of unitaries \cite{cleve2016efficient}. Quantum
algorithms also use stochastic trajectories \cite{li2023succinct} or
variational states \cite{endo2020variational} to simulate open-system
evolution. For non-Markovian dynamics, an additional challenge arises:
the environment retains information about earlier system--bath
interactions that can influence the subsequent system evolution.
Capturing this environmental memory requires a simulator to retain
enough bath degrees of freedom to reproduce their continuing influence
on the system. The central challenge is to identify which environmental
degrees of freedom must be represented, and how to encode them with
limited quantum resources.

Existing methods account for environmental memory in several ways. One
route retains bath modes explicitly and evolves them together with the
system, as in pseudomode methods \cite{tamascelli2018nonperturbative},
reaction-coordinate mappings \cite{ilessmith2014environmental}, and
chain mappings \cite{chin2011chain,prior2010efficient}. Another route
introduces auxiliary operators coupled to the system density matrix to
account for bath correlations, as in hierarchical equations of motion
\cite{tanimura1989time} and dissipaton formulations
\cite{li2024towards}. Generalized master equations incorporate the
system's past evolution through a memory kernel
\cite{wang2023simulating}, while non-Markovian stochastic Schr\"odinger
equations use temporally correlated noise together with
history-dependent terms \cite{li2023succinct}.

Here, we follow the explicit-bath route and construct a dynamically
equivalent environment whose joint evolution with the system can be
simulated on a quantum computer. The starting point is that the system
responds to the bath through its fluctuations and their correlations
across time. For a system initially uncorrelated with a linearly
coupled, centered Gaussian environment, these two-time correlations
completely determine the bath's influence on the reduced dynamics by
Wick's theorem. We can therefore replace the microscopic environment
with an effective bath, represented by auxiliary bosonic modes, that
reproduces the same correlation kernel. Over a finite simulation
interval, this kernel may be dominated by only a few temporal
correlation patterns. Our method identifies these patterns through its
spectral decomposition and represents each retained pattern by an
effective bosonic mode with a time-dependent system-bath coupling. When
the kernel spectrum decays rapidly, a small number of modes can retain
the relevant environmental memory, with a controlled error in the
reduced dynamics.

A compact effective bath, however, does not necessarily admit a compact
qubit encoding. Each bosonic mode can hold arbitrarily many excitations,
whereas a quantum register can represent only a finite number of
occupation levels. When the integrated mode couplings are moderate,
keeping a few levels may suffice. Stronger couplings can populate higher
levels and make this encoding costly. To address this limitation, we
introduce \emph{coupling dilution}: each effective mode is replaced by
several copies, with the coupling to each copy reduced by the square
root of the number of copies. Together, these copies preserve the
correlation kernel of the retained effective bath and, before truncating
their occupation levels, generate the same reduced dynamics as that
bath. The excitations are distributed among more modes, suppressing the
repeated excitation of any individual copy that causes occupation-cutoff
errors. Dilution thus allows the simulator to retain the collective
strong-coupling dynamics while using a fixed number of occupation levels
per mode, at the cost of additional bath registers.

We combine these constructions with qubit encodings and digital
Hamiltonian evolution to obtain a quantum simulation algorithm with
explicit error and resource bounds. The analysis connects the required
resources to three properties of the environment: how many temporal
correlation patterns are needed to describe its influence, how large
the accumulated mode couplings are, and how rapidly those couplings
vary. Rapid spectral
decay favors a small effective bath, while coupling dilution provides
further flexibility by trading additional bath registers for fewer
occupation levels per register. This tradeoff may be useful for
neutral-atom quantum processors, where large qubit arrays and
programmable connectivity offer resources for representing the effective
bath \cite{bluvstein2022quantum,evered2023high}. The resulting framework
offers a controlled route to representing environmental memory on a
quantum computer, with a tunable tradeoff between the number of bath
registers and their local dimension. The practical benefit depends on
the required circuit depth, circuit width and the implementation of the
time-dependent couplings, which our resource estimates help assess.

The rest of the paper is organized as follows. \Cref{sec:gaussian-model}
formulates the open-system model and establishes the equivalence of
Gaussian baths with the same two-time correlation kernel.
\Cref{sec:auxiliary-bath} constructs the finite-rank effective bath from
the spectral decomposition of the kernel and analyzes both the direct
occupation cutoff and the diluted construction.
\Cref{sec:digital-simulation} develops the qubit encoding and the
Trotter simulation formula, together with its simulation-error bound.
\Cref{sec:algorithm-resources} combines the error bounds into sufficient
parameter choices at a prescribed accuracy, derives the circuit width
and depth, and relates their cost to the bath properties.
Finally, \cref{sec:discussion}
gives further discussion, summarizes the main conclusions with future
work. The appendices provide supporting technical details.

\section{Gaussian Baths and Reduced System Dynamics}
\label{sec:gaussian-model}

We consider a system coupled to a bosonic bath through the
time-independent Hamiltonian
\begin{equation} \label{eq:system-bath-hamiltonian}
    H
    = H_s\otimes I_b+I_s\otimes H_b+S\otimes B,
\end{equation}
where $H_s$ and $H_b$ are the free system and bath Hamiltonians, and
$S$ and $B$ are Hermitian coupling operators. The system and bath are
initially uncorrelated, $\rho(0)=\rho_s(0)\otimes\rho_B$.

For simulation, we work in the interaction picture with respect to the
free bath Hamiltonian alone. The joint evolution is then generated by
\begin{equation} \label{eq:bath-interaction-hamiltonian}
    H_{bI}(t)=H_s\otimes I_b+S\otimes B(t),
\end{equation}
where $B(t):=\re^{\ri H_b t}B\re^{-\ri H_b t}$ is the bath
coupling operator evolved under the free bath Hamiltonian.
The free bath evolution is absorbed into $B(t)$, while $H_s$ remains
explicit. Since this picture transformation acts only on the bath, it
leaves the reduced system state unchanged. Tracing out the bath in this
picture therefore gives the Schr\"odinger-picture system state.

We assume that $\rho_B$ is Gaussian and that $B(t)$ is linear in the
bosonic creation and annihilation operators, with zero mean, $\tr[\rho_B
B(t)]=0$. These assumptions hold, for example, for a harmonic bath in
thermal equilibrium with linear coupling to the system. We refer to
baths satisfying these assumptions simply as \emph{Gaussian baths}
below. The zero mean removes the average bath force, leaving only its
fluctuations.

The relevant bath information is the two-time bath correlation function
\begin{equation} \label{eq:bath-correlation-kernel}
    G(t,s):=\tr\bigl[\rho_B B(t)B(s)\bigr].
\end{equation}
This function describes how bath fluctuations at two times are
correlated and carries information about the bath's memory. Since bath
operators at different times generally do not commute, $G(t,s)$ can be
complex and need not equal $G(s,t)$. We retain the full complex
correlation function and do not assume that it depends only on the time
difference.

For a general bath, higher correlation functions supply additional
information about its influence on the system. For a Gaussian bath,
Wick's theorem fixes all of them in terms of $G$: odd moments vanish,
and even moments are sums of products of two-time correlations.
Consequently, two Gaussian baths with the same $G(t,s)$ on $[0,T]$
generate the same reduced dynamics on that interval, provided the system
Hamiltonian, coupling operator, and initial system state are the same.
This is the standard Gaussian-bath equivalence principle: matching the
first moments and two-time correlations preserves the reduced dynamics
\cite{tamascelli2018nonperturbative}. Since the bath operators have zero
mean, matching $G$ alone is sufficient, without any weak-coupling or
Markov approximation. We state and derive this result in detail in
\cref{app:kernel-equivalence}.

We can therefore replace the original environment by an effective
Gaussian bath with the same correlation function, even if its modes and
Hamiltonian are different. In the next section, we use this freedom to
construct such a bath directly from $G$, with all auxiliary modes
initially in the vacuum state.

\section{Effective-Bath Construction}
\label{sec:auxiliary-bath}

In this section, we construct an effective bath suitable for quantum
simulation in three steps. We first realize the correlation function
$G$ exactly using bosonic modes initially in the vacuum state. We then
retain finitely many modes and approximate each retained oscillator by
a finite-dimensional system. For the latter step, we consider a direct
occupation cutoff and a diluted construction that reduces the coupling
per mode. We quantify the errors introduced by both approximations.

\subsection{From Bath Correlations to Effective Modes}

The spectral decomposition of $G$ identifies the temporal profiles that
an effective bath must reproduce. On the simulation interval $[0,T]$,
the correlation kernel defines a positive Hermitian integral operator,
whose spectral decomposition is
\begin{equation} \label{eq:bath-spectral-decomp}
    G(t,s)=\sum_{n=1}^{\infty}\lambda_n u_n(t)u_n^*(s),
    \qquad \lambda_1\geq\lambda_2\geq\cdots\geq0.
\end{equation}
Here the eigenfunctions $u_n$ are orthonormal on $[0,T]$. Each describes
a temporal profile, and its eigenvalue $\lambda_n$ measures its weight
in the bath correlations.

We associate each profile with an independent auxiliary oscillator,
initially in its vacuum state, and choose its coupling coefficient as
$g_n(t)=\sqrt{\lambda_n}\,u_n(t)$. The resulting bath operator is
\begin{equation} \label{eq:auxiliary-bath}
    A(t)=\sum_{n=1}^{\infty}
    \left(g_n(t)a_n+g_n^*(t)a_n^\dagger\right).
\end{equation}
The operators $a_n$ and $a_n^\dagger$ lower and raise the occupation of
mode $n$, while their conjugate coefficients make $A(t)$ Hermitian.
Write $\rho_A=\ket{\bzero}\bra{\bzero}$ for their joint vacuum state,
where $\ket{\bzero}=\bigotimes_n\ket{0}_n$.

The construction works because independent vacuum modes satisfy
$\bra{\bzero}a_n a_m^\dagger\ket{\bzero}=\delta_{nm}$, while the
other second moments vanish. Their contributions consequently add to
\begin{equation*}
    \tr[\rho_A A(t)A(s)]
    =\sum_{n=1}^{\infty}g_n(t)g_n^*(s)=G(t,s).
\end{equation*}
Since the vacuum is Gaussian and $A(t)$ is linear with zero mean, this
bath reproduces the original reduced dynamics exactly. Even when the
original bath is thermal, the auxiliary modes can start in the vacuum
state, because the temperature dependence of the bath correlation
function is encoded in their coupling coefficients $g_n(t)$.

For simulation, we retain the $N$ modes with the largest eigenvalues:
\begin{equation}\label{eq:retained-bath}
    A_N(t)=\sum_{n=1}^{N}
    \left(g_n(t)a_n+g_n^*(t)a_n^\dagger\right).
\end{equation}
Its correlation function $G_N$ is the first $N$ terms of the spectral
decomposition \eqref{eq:bath-spectral-decomp}. This is the first
approximation in the construction: the discarded modes contribute the
correlation tail $G-G_N$. To connect this tail to simulation accuracy,
let $\rho_s(T)$ and $\rho_s^{(N)}(T)$ denote the reduced states for the
original and retained baths, with the same system Hamiltonian, coupling,
and initial system state. Their trace-norm difference obeys
\begin{equation} \label{eq:rank-truncation-bound}
    \tracenorm{\rho_s(T)-\rho_s^{(N)}(T)}
    \leq \exp\!\left(2\opnorm{S}^2\eta_N\right)-1,
    \qquad
    \eta_N:=T\left(\sum_{n>N}\lambda_n^2\right)^{1/2}.
\end{equation}
The precise statement and proof appear in
\cref{app:rank-truncation-error}. Thus we increase $N$ until this bound
falls below the allocated mode-truncation error $\veps_{\rank}$. Rapid
eigenvalue decay allows a small number of modes to reproduce the bath
correlations accurately.

The required rank reflects the temporal structure of the bath
correlations rather than the microscopic size of the environment. A
single-frequency correlation $G(t,s)=A\re^{-\ri\omega_0(t-s)}$ has rank
one, so one effective mode suffices regardless of how many microscopic
oscillators contribute at that frequency. More generally, kernel
smoothness provides a guide to spectral compression: greater regularity
gives faster eigenvalue decay, since $G\in C^p([0,T]^2)$ implies
$\lambda_n=o(n^{-(p+1)})$ \cite{reade1984eigenvalues}. For example, the
zero-temperature Ohmic correlation with exponential cutoff,
$G(t,s)=\eta\omega_c^2/[1+\ri\omega_c(t-s)]^2$, is analytic, allowing
efficient spectral compression on a fixed simulation interval.

In practice, the required number of modes depends on the correlation
function, the simulation interval, and the desired accuracy. Even after
this truncation, each retained mode is still an infinite-dimensional
oscillator. The next subsection addresses this remaining cost by
restricting the local occupations.

\subsection{Finite occupation cutoff}
\label{subsec:nondiluted-d-level}

The retained bath has only $N$ modes, but each oscillator can still hold
arbitrarily many excitations. To represent it with a finite quantum
register, we keep only the occupation levels $0,\ldots,d-1$ of each
mode, with $d\geq2$. This approximation removes transitions to higher
levels and is useful when those levels contribute little to the system's
evolution. We call the resulting $d$-level system a \emph{finite-level
mode}, whose local occupation space is
$\operatorname{span}\{\ket{0},\ldots,\ket{d-1}\}$. When implementing the
simulation, each finite-level mode is encoded in a quantum register, which
is called a \emph{bath register}.

The truncated annihilation operator in the Fock basis
$\{\ket{j}\}_{j=0}^{\infty}$ is
\begin{equation}\label{eq:truncated-annihilation-operator}
    a_n^{(d)}
    =\sum_{j=1}^{d-1}\sqrt{j}\,
    \left(\ket{j-1}\bra{j}\right)_n,
\end{equation}
and its adjoint is the truncated creation operator. Replacing the
bosonic operators in $A_N(t)$ gives the finite-level bath field
\begin{equation*}
    A_{N,d}(t)
    =\sum_{n=1}^{N}
    \left(g_n(t)a_n^{(d)}
    +g_n^*(t)(a_n^{(d)})^\dagger\right).
\end{equation*}
All finite-level modes start in the vacuum state $\ket{0}$. Since the
vacuum two-time correlation function uses only the transition between
$\ket{0}$ and $\ket{1}$, the cutoff preserves the retained correlation
kernel:
\begin{equation*}
    \bra{\bzero}A_{N,d}(t)A_{N,d}(s)\ket{\bzero}=G_N(t,s).
\end{equation*}
However, the finite-level bath is no longer Gaussian, so matching this
kernel does not guarantee the same reduced dynamics. The equivalence
principle of \cref{sec:gaussian-model} therefore no longer applies.
The error comes from higher-order transitions that would populate the
omitted occupation levels.

To quantify this error, define the accumulated coupling of mode $n$ by
\begin{equation*}
    \beta_{n,t}:=\opnorm{S}\int_0^t|g_n(u)|\,\rd u.
\end{equation*}
This quantity accounts for both the coupling strength and the duration
of the interaction. Let $\rho_s^{(N,d)}(T)$ denote the reduced state
obtained with the finite-level bath, with the same system Hamiltonian,
coupling operator, and initial system state as for the retained bosonic
bath. The occupation-cutoff error obeys
\begin{equation}\label{eq:direct-cutoff-error}
    \tracenorm{\rho_s^{(N)}(T)-\rho_s^{(N,d)}(T)}
    \leq\frac{2^d}{\sqrt{d!}}\sum_{n=1}^{N}\beta_{n,T}^d.
\end{equation}
The precise statement and proof are given in
\cref{app:occupation-cutoff-error}. We choose $d$ so that this
bound lies below the allocated occupation-cutoff error
$\veps_{\mathrm{occ}}$.

For fixed accumulated couplings, the factorial in the denominator
ensures convergence as more levels are retained. However, it competes
with the factor $\beta_{n,T}^d$, so a small cutoff may not be sufficient
when the accumulated coupling is large.
For example, for a single retained mode, the bound guarantees an error
below $10^{-2}$ with $d=17$ when $\beta_T=1$, whereas achieving the
same guarantee when $\beta_T=10$ requires $d=1093$. Although these
sufficient cutoffs may be conservative, they illustrate how the local
dimension needed for this error guarantee grows with accumulated coupling.

A physical illustration is a vacuum oscillator driven by
$H_{\mathrm{drv}}(t)=f(t)(a+a^\dagger)$, with real, nonnegative $f(t)$
and accumulated drive $\theta:=\int_0^T f(t)\,\rd t$. Since the
Hamiltonians commute at different times, the final state is
\begin{equation*}
    \ket{\psi(T)}
    =\re^{-\ri\theta(a+a^\dagger)}\ket0
    =\re^{-\theta^2/2}
      \sum_{k=0}^{\infty}
      \frac{(-\ri\theta)^k}{\sqrt{k!}}\ket k.
\end{equation*}
This coherent state has a Poisson occupation distribution with mean
$\bar n=\theta^2$. Increasing the accumulated drive from $\theta=1$
to $\theta=10$ therefore increases the mean occupation from $1$ to
$100$, shifting the distribution toward higher levels. A cutoff that
retains only a few levels then captures little of the occupation
probability. This example provides physical intuition for how a large
accumulated drive can populate high occupation levels, making a small
local cutoff inadequate for representing the oscillator state.

A larger local dimension also increases the implementation burden on a
quantum computer, since more occupation levels and transitions must be
represented and simulated. This motivates distributing the bath
excitations among more modes, each retaining fewer occupation levels.

\subsection{Coupling dilution}
\label{subsec:coupling-dilution}

Coupling dilution avoids the large local occupation by replacing each
retained mode with several copies, each coupled more weakly to the
system. We choose the couplings so that the copies together reproduce
the original mode's contribution to the bath correlation kernel. Before
imposing a local occupation cutoff, this replacement therefore preserves
the reduced system dynamics exactly. The smaller coupling to each copy
then allows us to control the cutoff error while retaining fewer levels
per mode.

For each retained mode $n$, introduce $q$ independent bosonic modes
$a_{n,r}$, $r=1,\ldots,q$, initially in vacuum, and reduce each coupling
by a factor $\sqrt q$. The resulting bath field is
\begin{equation*}
    A_{N,q}(t)
    =\sum_{n=1}^{N}\sum_{r=1}^{q}
    \left(\frac{g_n(t)}{\sqrt q}a_{n,r}
    +\frac{g_n^*(t)}{\sqrt q}a_{n,r}^\dagger\right).
\end{equation*}
Here $q\geq1$ is an integer, called the dilution factor. Independent
vacuum modes satisfy
$\bra{\bzero}a_{n,r}a_{m,r'}^\dagger\ket{\bzero}
=\delta_{nm}\delta_{rr'}$, while the other second moments vanish.
The diluted bath therefore has correlation kernel
\begin{equation*}
    \bra{\bzero}A_{N,q}(t)A_{N,q}(s)\ket{\bzero}
    =\sum_{n=1}^{N}\sum_{r=1}^{q}\frac{g_n(t)g_n^*(s)}{q}
    =G_N(t,s).
\end{equation*}
The $q$ copies together retain the original correlation strength,
although each couples more weakly to the system. Since this bath is
still Gaussian and centered, the equivalence principle in 
\cref{sec:gaussian-model} guarantees the same reduced dynamics as
the retained bosonic bath $A_N(t)$.

For simulation, we replace each bosonic copy by a finite-level mode
with occupation levels $0,\ldots,d-1$.
Using the truncated operators defined in
\eqref{eq:truncated-annihilation-operator} for each copy gives
\begin{equation}\label{eq:diluted-finite-level-bath}
    A_{N,q,d}(t)
    =\sum_{n=1}^{N}\sum_{r=1}^{q}
    \left(\frac{g_n(t)}{\sqrt q}a_{n,r}^{(d)}
    +\frac{g_n^*(t)}{\sqrt q}(a_{n,r}^{(d)})^\dagger\right).
\end{equation}
As with the direct cutoff, the finite-level bath preserves $G_N$ for
$d\geq2$ but is no longer Gaussian. Its error comes from transitions that
would excite a bosonic copy beyond the retained levels. 

Let $\rho_s^{(N,q,d)}(T)$ denote the reduced state generated by this
finite-level bath, with the same system Hamiltonian, coupling operator,
and initial system state as before. Its occupation-cutoff error obeys
\begin{equation}\label{eq:diluted-cutoff-error}
    \tracenorm{\rho_s^{(N)}(T)-\rho_s^{(N,q,d)}(T)}
    \leq\frac{2^d}{q^{d-1}}\sum_{n=1}^{N}\beta_{n,T}^{2d}.
\end{equation}
The estimate above uses additional cancellation under the bath trace and
does not follow from simply applying \eqref{eq:direct-cutoff-error} to
the diluted couplings. In \cref{app:occupation-cutoff-error}, we
establish two complementary upper bounds: a factorial bound suited to
increasing the local dimension, and the bound used here, suited to
increasing the number of copies. For fixed $d\geq2$, the bound decreases
as $q^{1-d}$, so we can meet the occupation-error budget
$\veps_{\mathrm{occ}}$ by increasing the number of copies. For example,
$d=2$ gives an error bound $4\sum_n\beta_{n,T}^4/q$ using a two-level
finite-level mode for each copy.

The two constructions therefore offer a choice: increase the number of
levels in each of the $N$ finite-level modes, or use $Nq$ finite-level
modes with fewer levels each. Dilution can accommodate larger
accumulated couplings at fixed local dimension, at the cost of more
finite-level modes. The overall resource cost depends on both their
encoding into bath registers and the evolution circuit, which we address
next.

\section{Digital Quantum Simulation}
\label{sec:digital-simulation}

The construction in \cref{sec:auxiliary-bath} replaces the original bath
by a collection of finite-level modes, with controlled errors from rank
truncation and the occupation cutoff. We now describe how to simulate
the joint evolution of the system and these finite-level modes on a
quantum computer. Replacing the coupling operator $B(t)$ in the bath
interaction-picture Hamiltonian \eqref{eq:bath-interaction-hamiltonian}
with the finite-level bath field $A_{N,q,d}(t)$, we obtain the simulator
Hamiltonian
\begin{equation*}
    H_{N,q,d}(t)
    =H_s\otimes I_b+S\otimes A_{N,q,d}(t),
\end{equation*}
where $q=1$ includes the direct occupation cutoff. Each finite-level
mode starts in $\ket{0}$, and the time-dependent couplings encode the
retained bath correlations.

We implement this evolution using a second-order Trotter formula that
exploits the commuting structure of the interactions with the
finite-level modes. We first identify the commuting transitions, then
construct the approximate unitary evolution and bound its error.
Throughout this section, $N,q,d$ are fixed, and we write
$H(t)=H_{N,q,d}(t)$ and $U(T)=U_{N,q,d}(T)$ when the parameters need not
be displayed.

\subsection{Hamiltonian decomposition}

We first separate the system Hamiltonian from its interaction with the
finite bath:
\begin{equation} \label{eq:finite-bath-mode-split}
    H(t)=H_0+H_{\mathrm{int}}(t),
    \qquad 
    H_0:=H_s\otimes I_b,
    \quad 
    H_{\mathrm{int}}(t):=S\otimes A_{N,q,d}(t).
\end{equation}
The first term acts only on the system, while the second couples the
system to the finite-level bath modes. To decompose this interaction,
recall that each truncated ladder operator connects neighboring
occupation levels. Label a finite-level mode by $(n,r)$, where $n$
identifies the retained correlation profile and $r$ labels its copy. The
transition between levels $j-1$ and $j$ of this mode is generated by
\begin{equation} \label{eq:finite-bath-transition}
    H_{n,r,j}(t)
    :=\frac{\sqrt j}{\sqrt q}\,S\otimes
    \left(g_n(t)\ket{j-1}\bra j
    +g_n^*(t)\ket j\bra{j-1}\right)_{n,r}.
\end{equation}
Expanding $A_{N,q,d}(t)$ in these transitions gives
\begin{equation*}
    H_{\mathrm{int}}(t)
    =\sum_{n=1}^{N}\sum_{r=1}^{q}\sum_{j=1}^{d-1}H_{n,r,j}(t).
\end{equation*}

Adjacent transitions share an occupation level and generally do not
commute. We therefore divide the ladder into alternating links: odd
transitions connect $(0,1),(2,3),\ldots$, and even transitions connect
$(1,2),(3,4),\ldots$, stopping at level $d-1$. Within each group the
pairs are disjoint, so the transition Hamiltonians commute. Transitions
on different modes also commute, since their bath operators act on
different registers, and all couple through the same system operator
$S$.

Let $\sfJ_p$ denote the transition indices in group $p$ for $p \in
\{\even, \odd\}$:
\begin{equation*}
    \sfJ_{\even}:=\{1\leq j\leq d-1:j\text{ is even}\},
    \qquad
    \sfJ_{\odd}:=\{1\leq j\leq d-1:j\text{ is odd}\}.
\end{equation*}
The corresponding Hamiltonians for one mode and for the whole bath are
\begin{equation*}
    H_{n,r,p}(t):=\sum_{j\in\sfJ_p}H_{n,r,j}(t),
    \qquad
    H_p(t):=\sum_{n=1}^{N}\sum_{r=1}^{q}H_{n,r,p}(t),
    \quad p\in\{\even,\odd\}.
\end{equation*}
This gives the decomposition used below,
\begin{equation} \label{eq:finite-bath-parity-split}
    H(t)=H_0+H_{\mathrm{int}}(t)
    =H_0+H_{\even}(t)+H_{\odd}(t).
\end{equation}
Each parity Hamiltonian is a sum of commuting transitions. The two
parity Hamiltonians generally do not commute with each other, nor does
$H_0$ generally commute with the interaction. These are the two
splittings needed for the evolution circuit.

\subsection{Unitary evolution and circuit implementation}

The exact evolution of the system and finite bath is
\begin{equation*}
    U(T)\equiv U_{N,q,d}(T)
    =\calT\exp\left(-\ri\int_0^T H(t)\,\rd t\right).
\end{equation*}
To approximate it, divide $[0,T]$ into $M_{\trot}$ intervals of length
$\Delta t=T/M_{\trot}$, with $t_k=k\Delta t$ and midpoint
$m_k=t_k+\Delta t/2$. On each interval, evaluate the time-dependent
interaction Hamiltonian at $m_k$ and use a symmetric Strang splitting
between the system and interaction terms:
\begin{equation*}
    \re^{-\ri \Delta t(H_0+H_{\mathrm{int}}(m_k))}
    \approx
    \re^{-\ri \Delta t H_0/2}
    \re^{-\ri \Delta t H_{\mathrm{int}}(m_k)}
    \re^{-\ri \Delta t H_0/2}.
\end{equation*}
The middle factor still contains overlapping occupation transitions.
We separate it into even and odd operations with a second-order Strang
splitting:
\begin{equation*}
    \re^{-\ri \Delta t H_{\mathrm{int}}(m_k)}
    \approx
    \re^{-\ri \Delta t H_{\even}(m_k)/2}
    \re^{-\ri \Delta t H_{\odd}(m_k)}
    \re^{-\ri \Delta t H_{\even}(m_k)/2}.
\end{equation*}
We place the even transitions in the two half steps because this group
contains fewer transitions when $d$ is even, reducing the number of
transition operations per step. When $d$ is odd, the two groups have
equal size, so either ordering has the same cost.

Combining the two splittings gives the single-step evolution
\begin{equation} \label{eq:fully-decomposed-step}
    W_k:=
    \re^{-\ri \Delta t H_0/2} \,
    \re^{-\ri \Delta t H_{\even}(m_k)/2} \,
    \re^{-\ri \Delta t H_{\odd}(m_k)} \,
    \re^{-\ri \Delta t H_{\even}(m_k)/2} \,
    \re^{-\ri \Delta t H_0/2}.
\end{equation}
Each parity factor can now be decomposed exactly into elementary
transition operations:
\begin{equation*}
    \re^{-\ri\tau H_p(t)}
    =\prod_{n=1}^{N}\prod_{r=1}^{q}\prod_{j\in\sfJ_p}
    \re^{-\ri\tau H_{n,r,j}(t)},
    \qquad p\in\{\even,\odd\}.
\end{equation*}

\begin{figure}[htbp]
    \centering
    \begin{tikzpicture}[
        x=0.86cm,y=0.55cm,font=\scriptsize,
        wire/.style={draw=gray!65,thick},
        sys/.style={draw=blue!70!black,fill=blue!8,
            rounded corners=1.5pt,thick,minimum width=0.70cm,
            minimum height=0.48cm,inner sep=1pt},
        even/.style={draw=cyan!55!black,fill=cyan!10,
            rounded corners=1.5pt,thick,minimum width=0.64cm,
            minimum height=0.44cm,inner sep=1pt},
        odd/.style={draw=orange!80!black,fill=orange!12,
            rounded corners=1.5pt,thick,minimum width=0.64cm,
            minimum height=0.44cm,inner sep=1pt}
    ]
        \foreach \y in {0,-1.4,-2.6,-4.2}
            \draw[wire] (0,\y)--(15.4,\y);
        \node[anchor=east] at (-0.12,0) {system};
        \node[anchor=east] at (-0.12,-1.4) {$\mathrm{reg}_1$};
        \node[anchor=east] at (-0.12,-2.6) {$\mathrm{reg}_2$};
        \node[anchor=east] at (-0.12,-3.3) {$\vdots$};
        \node[anchor=east] at (-0.12,-4.2) {$\mathrm{reg}_{Nq}$};
        \node[sys] at (0.8,0) {$U_0^{1/2}$};
        \node[sys] at (14.6,0) {$U_0^{1/2}$};
        \foreach \x/\y/\lab in {
            2.0/-1.4/1,3.0/-2.6/2,4.8/-4.2/{Nq},
            10.6/-1.4/1,11.6/-2.6/2,13.4/-4.2/{Nq}
        } {
            \draw[cyan!55!black,thick] (\x,0)--(\x,\y);
            \node[even] at (\x,0) {$E_{\lab}$};
            \node[even] at (\x,\y) {$E_{\lab}$};
        }
        \foreach \x/\y/\lab in {
            6.3/-1.4/1,7.3/-2.6/2,9.1/-4.2/{Nq}
        } {
            \draw[orange!80!black,thick] (\x,0)--(\x,\y);
            \node[odd] at (\x,0) {$O_{\lab}$};
            \node[odd] at (\x,\y) {$O_{\lab}$};
        }
        \foreach \x in {3.9,8.2,12.5} {
            \node[fill=white,inner sep=1pt] at (\x,0) {$\cdots$};
            \node[fill=white,inner sep=1pt,text=gray!70]
                at (\x,-3.3) {$\ddots$};
        }
        \node[text=cyan!55!black] at (3.4,0.95) {even half step};
        \node[text=orange!80!black] at (7.7,0.95) {odd full step};
        \node[text=cyan!55!black] at (12.0,0.95) {even half step};
    \end{tikzpicture}
    \caption{Circuit of one midpoint--Strang step
    $W_k$, read from left to right. Each bath wire represents a register
    encoding one finite-level mode. Registers $1$, $2$, and $Nq$ are
    shown, with intermediate registers omitted; the single index is
    $\ell=(n-1)q+r$.
    Each connected pair of boxes denotes one joint operation on the
    system and the indicated bath register. The cyan $E_{\ell}$ blocks
    contain the even transitions for a half step, and the orange
    $O_{\ell}$ blocks contain the odd transitions for a full step.
    Blocks within each parity group commute, but share the system
    register, so parallel execution depends on the hardware architecture.}
    \label{fig:fully-decomposed-slice}
\end{figure}
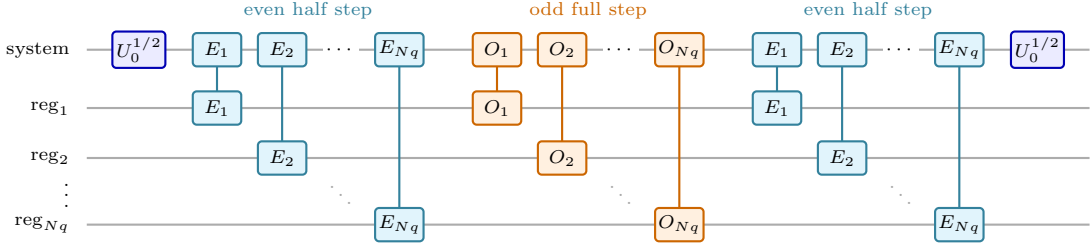

\Cref{fig:fully-decomposed-slice} shows the circuit for one time step,
with transition operations grouped by bath register. For the figure,
we number the registers by $\ell=(n-1)q+r$ and write
\begin{equation*}
    U_0^{1/2}:=\re^{-\ri \Delta t H_0/2},
    \qquad
    E_{\ell}:=\re^{-\ri \Delta t H_{n,r,\even}(m_k)/2},
    \qquad
    O_{\ell}:=\re^{-\ri \Delta t H_{n,r,\odd}(m_k)}.
\end{equation*}
Each $E_{\ell}$ or $O_{\ell}$ is a product of elementary transition
operations of the corresponding parity, with all couplings evaluated
at the same midpoint $m_k$.

The complete digital evolution is
\begin{equation*}
    \widehat U(T)\equiv\widehat U_{N,q,d}(T)=W_{M_{\trot}-1}\cdots W_1W_0,
\end{equation*}
with later steps acting to the left. Since $H_0$ is time independent,
neighboring system half steps can be merged. When $d=2$, the even blocks
disappear and each local odd block contains only the transition between
$\ket{0}$ and $\ket{1}$.
The inner splitting is then exact; midpoint freezing and the outer
system--interaction splitting still contribute to the simulation error.

\subsection{Digital simulation error}
\label{subsec:digital-error}

The digital evolution introduces three approximations: freezing the
couplings at the midpoint of each time interval, separating the system
and interaction evolution, and separating the even and odd transition
evolution. The first error depends on how rapidly the couplings vary,
while the latter two depend on the noncommutativity of the separated
Hamiltonian terms. We assume exact implementation of the system and
elementary transition unitaries. Their gate decompositions depend on the
chosen encoding and hardware. Any further gate-synthesis error should be
accounted for separately.

Let $\rho_s^{(N,q,d)}(T)$ be the exact reduced state for the finite bath
and $\widehat\rho_s(T)$ the state obtained with the circuit
\eqref{eq:fully-decomposed-step}, starting from the same system state
and vacuum bath registers. Assume that $H_{\mathrm{int}}(t)$ is twice
continuously differentiable in operator norm on $[0,T]$. For $M_{\trot}$
uniform steps of duration $\Delta t=T/M_{\trot}$, the digital error
obeys
\begin{equation}\label{eq:digital-simulation-error}
    \tracenorm{\rho_s^{(N,q,d)}(T)-\widehat\rho_s(T)}
    \leq T(\Delta t)^2R_{\trot}.
\end{equation}
The coefficient $R_{\trot}=R_{\mathrm{mid}}+R_{\mathrm{out}}+
R_{\mathrm{in}}$ collects the contributions from midpoint evaluation,
the outer system--interaction splitting, and the inner even--odd
splitting, respectively. The first two contributions are independent of
the local cutoff $d$ and dilution factor $q$. The inner splitting
contribution scales as $(d-1)^{3/2}q^{-1/2}$ for $d\geq3$ and vanishes
for $d=2$. Thus, for fixed $N,d,T$ and retained couplings, increasing
$q$ can reduce the number of time steps required to meet a prescribed
digital-error tolerance under this bound; for $d=2$, the bound is
independent of $q$. The explicit expressions for these contributions,
the formal statement, and the proof are given in
\cref{app:trotterization-error}.

Coupling dilution can be particularly useful when the inner even--odd
splitting contribution dominates the digital-error bound. For fixed
$N,T$, retained couplings, and local dimension $d\geq3$, this
contribution decreases as $q^{-1/2}$, giving a sufficient time-step
count proportional to $q^{-1/4}$ while it remains dominant. Thus
dilution can reduce both the occupation-cutoff bound and the required
number of time steps in this regime. When $d=2$, each bath register has
only the transition between $\ket{0}$ and $\ket{1}$, so the
even-transition Hamiltonian vanishes. The inner even--odd splitting is
therefore exact and contributes no error to $R_{\trot}$. This simplifies
the circuit and removes one contribution to the error bound, although
the occupation cutoff may require more diluted copies to achieve the
desired accuracy.

\section{Simulation Algorithm and Resource Analysis}
\label{sec:algorithm-resources}

The effective-bath construction and digital evolution described above
provide the ingredients of our simulation algorithm. We now combine
them into a procedure for approximating the reduced system state at a
prescribed accuracy. We first allocate the error budget and select the
simulation parameters, then describe the numerical construction of the
coupling functions and summarize the algorithm. We finally estimate the
resulting circuit width and depth.

\subsection{Algorithm and error control}

For a simulation time $T$ and target trace-norm error $0<\veps<1$,
allocate positive budgets for the rank truncation, occupation cutoff and
digital error to achieve
\begin{equation}\label{eq:total-simulation-error}
    \tracenorm{\rho_s(T)-\widehat\rho_s(T)}
    \leq
    \veps_{\rank}+\veps_{\mathrm{occ}}+\veps_{\trot}
    \leq\veps,
\end{equation}
provided the parameters meet the corresponding bounds below. We first
choose the retained rank $N$, then the local dimension $d$ and dilution
factor $q$, and finally the number of evolution steps $M_{\trot}$.
This guarantee assumes exact coupling functions and elementary unitaries;
numerical mode reconstruction and gate synthesis require additional
error allowances within the target $\veps$.

\paragraph{Parameter selection.}
The following prescriptions are sufficient rather than optimal; smaller
values may already satisfy the original error bounds.

We first choose how many effective temporal modes to retain. By the
rank-truncation bound in \eqref{eq:rank-truncation-bound}, the discarded
modes contribute at most $\exp(2\opnorm{S}^2\eta_N)-1$ to the
reduced-state error, where $\eta_N=T(\sum_{n>N}\lambda_n^2)^{1/2}$ is
the accumulated spectral tail bound. Requiring this contribution to be
at most $\veps_{\rank}$ gives the rank criterion
\begin{equation} \label{eq:rank-truncation-choice}
    T\left(\sum_{n>N}\lambda_n^2\right)^{1/2}
    \leq\frac{\ln(1+\veps_{\rank})}{2\opnorm{S}^2}.
\end{equation}

Next choose the occupation cutoff for the retained modes. Recall that
the occupation-error depends on the accumulated couplings
$\beta_{n,T}=\opnorm{S}\int_0^T|g_n(t)|\,\rd t$. For the direct-cutoff
construction, write $\beta_{\max}=\max_{1\leq n\leq N}\beta_{n,T}$.
Using $d!\geq(d/\re)^d$ in the direct-cutoff bound
\eqref{eq:direct-cutoff-error} gives
\begin{equation*}
    \tracenorm{\rho_s^{(N)}(T)-\rho_s^{(N,d)}(T)}
    \leq N\left(2\beta_{\max}\sqrt{\frac{\re}{d}}\right)^d.
\end{equation*}
Choosing $d\geq16\re\,\beta_{\max}^2$ makes the factor in parentheses at
most $1/2$. The error is then bounded by $N2^{-d}$, which is at most
$\veps_{\mathrm{occ}}$ when $d\geq\log_2(N/\veps_{\mathrm{occ}})$.
Combining these conditions with the minimum local dimension gives the
sufficient choice
\begin{equation*}
    d\geq\max\left\{
        2,\left\lceil16\re\,\beta_{\max}^2\right\rceil,
        \left\lceil\log_2\frac{N}{\veps_{\mathrm{occ}}}\right\rceil
    \right\}.
    \label{eq:resource-direct-choice}
\end{equation*}
Here, the second term, which scales with $\beta_{\max}^2$, accounts for
the strongest accumulated coupling, while the logarithmic term provides
sufficient suppression of the error summed over all retained modes.

Alternatively, fix an integer local dimension $d\geq2$ and distribute
each mode's coupling over $q$ copies. The diluted-cutoff bound
\eqref{eq:diluted-cutoff-error} requires
\begin{equation*}
    \frac{2^d}{q^{d-1}}\sum_{n=1}^{N}\beta_{n,T}^{2d}
    \leq\veps_{\mathrm{occ}}.
\end{equation*}
Solving for the integer dilution factor gives
\begin{equation}
    q\geq\max\left\{
        1,\left\lceil
        \left(\frac{2^d\sum_{n=1}^{N}\beta_{n,T}^{2d}}
        {\veps_{\mathrm{occ}}}\right)^{1/(d-1)}
        \right\rceil
    \right\}.
    \label{eq:resource-diluted-choice}
\end{equation}
For example, $d=2$ requires $q\geq4\sum_n\beta_{n,T}^4/
\veps_{\mathrm{occ}}$. Thus the occupation error can be reduced either
by retaining more levels in each mode or by using more weakly coupled
copies at a fixed local dimension. 

Finally, the second-order digital-error bound
\eqref{eq:digital-simulation-error} gives the sufficient step count
\begin{equation}\label{eq:resource-step-choice}
    M_{\trot}\geq
    \max\left\{1,\left\lceil
        \sqrt{\frac{R_{\trot}T^3}{\veps_{\trot}}}
    \right\rceil\right\}.
\end{equation}
The explicit definition and derivation of $R_{\trot}$ are given in
\cref{app:trotterization-error}. The step count is chosen after fixing
$N,d,q$ because these parameters also determine the Hamiltonian entering
$R_{\trot}$.

\paragraph{Construction of the coupling functions.}

To compute the coupling functions of the effective modes numerically,
discretize the kernel eigenvalue equation using the midpoint rule on
$[0,T]$. Divide $[0,T]$ into $M$ intervals of length $\Delta
t_{\mathrm{ker}}=T/M$. Taking $t_j=(j-\tfrac12)\Delta t_{\mathrm{ker}}$
gives
\begin{equation*}
    \Delta t_{\mathrm{ker}}\sum_{k=1}^{M}
    G(t_j,t_k)u_n(t_k)
    \approx \lambda_n u_n(t_j).
\end{equation*}
Thus, diagonalizing the matrix $K_{jk}=\Delta
t_{\mathrm{ker}}G(t_j,t_k)$ gives approximate eigenvalues $\hlambda_n$
and normalized eigenvectors $V_{jn}$. Since the continuum normalization
becomes $\Delta t_{\mathrm{ker}}\sum_j|u_n(t_j)|^2\approx1$, the sampled
eigenfunctions satisfy $u_n(t_j)\approx V_{jn}/\sqrt{\Delta
t_{\mathrm{ker}}}$. Retaining the leading $N$ eigenpairs gives the
sampled coupling functions
\begin{equation*}
    g_n(t_j)\approx
    \sqrt{\frac{\hlambda_n}{\Delta t_{\mathrm{ker}}}}\,V_{jn},
    \qquad n=1,\ldots,N,
\end{equation*}
which are interpolated for evaluation at the circuit's time
points if the quantum evolution uses a different time grid.
Grid convergence alone does not certify the continuum rank criterion.

\paragraph{Algorithm summary.}
We summarize the complete workflow, from classical mode reconstruction
and parameter selection to quantum evolution and recovery of the reduced
system state in \cref{alg:effective-bath-simulation}.

\begin{algorithm}[htbp]
    \caption{Quantum simulation with a finite effective bath}
    \label{alg:effective-bath-simulation}
    \algrenewcommand{\algorithmicrequire}{\textbf{Input:}}
    \algrenewcommand{\algorithmicensure}{\textbf{Output:}}
    \begin{algorithmic}[1]
        \Require Initial state $\rho_s(0)$, operators $H_s,S$, correlation
        kernel $G$, evolution time $T$, and target error $\veps$.
        \Ensure Approximate reduced system state $\widehat\rho_s(T)$.
        \State Allocate the error budgets
        $\veps_{\rank}+\veps_{\mathrm{occ}}+\veps_{\trot}\leq\veps$.
        \Statex \textbf{Construct the effective modes:}
        \State Diagonalize the discretized matrix $K$ of kernel $G$. 
        \State Retain the leading $N$ eigenpairs by rank criterion \eqref{eq:rank-truncation-choice}.
        \Statex \textbf{Choose the finite bath:}
        \State Compute accumulated coupling $\beta_{n,T}$.
        \If{using the direct cutoff}
            \State Set $q=1$ and choose $d$ by
            \eqref{eq:resource-direct-choice}.
        \Else
            \State Fix $d\geq2$ and choose $q$ by
            \eqref{eq:resource-diluted-choice}.
        \EndIf
        \Statex \textbf{Prepare and evolve:}
        \State Evaluate or bound $R_{\trot}$, choose $M_{\trot}$ by
        \eqref{eq:resource-step-choice}, and set $\Delta t=T/M_{\trot}$.
        \State Prepare the system in $\rho_s(0)$ and all $Nq$ bath
        registers in $\ket{0}$.
        \For{$k=0,\ldots,M_{\trot}-1$}
            \State Apply one-step evolution $W_k$ from \eqref{eq:fully-decomposed-step}.
        \EndFor
        \State Discard the bath registers.
        \State \Return $\widehat\rho_s(T)$, the resulting system state.
    \end{algorithmic}
\end{algorithm}

\begin{remark}[Mode-dependent occupation parameters]
    In practice, the local dimension and dilution factor can be chosen
    separately for each mode according to its accumulated coupling
    $\beta_{n,T}$. Weakly coupled modes may then use fewer occupation
    levels or copies, provided the sum of the modewise error bounds
    remains within the occupation-error budget.
\end{remark}

\subsection{Circuit resource estimates}
\label{subsec:circuit-resources}

We now show that the circuit width and depth are bounded polynomially in
the simulation time $T$ and inverse accuracy $1/\veps$, under the bath
assumptions and circuit implementation conditions stated below. We
establish this by combining the circuit cost for given $N,d,q$ with
sufficient parameter scalings from the error bounds. 

\paragraph{Width and depth.}
There are different schemes to map the $d$ occupation levels of each
finite-level mode into states of a qubit register
\cite{sawaya2020resource}. For example, binary encoding stores the
occupation number in $\lceil\log_2d\rceil$ qubits, while unary encoding
uses $d$ qubits, with one excitation marking the occupied level. Denote
the chosen encoding by $\mathsf e$ and its qubit count by $w_{\mathsf
e}(d)$. For $n_s$ system qubits and $Nq$ bath registers, the register
width is
\begin{equation}\label{eq:resource-width}
    W_{\mathrm{circ}}=n_s+Nq\,w_{\mathsf e}(d).
\end{equation}
The encoding also determines how occupation transitions are implemented
as qubit operations, and hence their circuit depth. Let
$D_{\mathrm{sys}}$ bound the depth of one system-evolution unitary and
$D_{\mathrm{int}}(d,\mathsf e)$ that of one elementary system--bath
transition unitary. Each step contains $O(Nqd)$ such transition
unitaries, so serial execution gives
\begin{equation}\label{eq:resource-depth}
    D_{\mathrm{circ}}=O\left[
        M_{\trot}\bigl(D_{\mathrm{sys}}
        +NqdD_{\mathrm{int}}(d,\mathsf e)\bigr)\right].
\end{equation}
The depths of these elementary unitaries depend on the system operators,
encoding, and required gate precision. Interactions sharing system
qubits are counted serially even when their generators commute.

\paragraph{Sufficient parameter scalings.}
For the scaling estimates, fix the system operators and consider
$T\geq1$ in fixed time units and $0<\veps<1$, with each error budget a
fixed positive fraction of $\veps$. We assume that the kernel
eigenvalues $\lambda_{n,T}$ on $[0,T]$ satisfy
\begin{equation*}
    \lambda_{n,T}\leq C_\lambda T^\kappa n^{-p},
    \qquad p>1.
\end{equation*}
This condition controls both the retained rank and occupation
parameters. Further, for time evolution, assume that the effective
couplings are twice continuously differentiable, with derivatives
satisfying
\begin{equation*}
    \max_{0\leq j\leq2}\sup_{n\geq1}
    \sup_{t\in[0,T]}|g_n^{(j)}(t)|
    \leq C_g T^c.
\end{equation*}
Here $\kappa,c\geq0$, and the constants are independent of $T$ and
$\veps$. These assumptions are sufficient for the scaling estimates.

For the retained rank, the spectral decay and rank-error bound allow
\begin{equation*}
    N=O(T^a\veps^{-b}),
    \qquad a:=\frac{\kappa+1}{p-1/2},
    \qquad b:=\frac{1}{p-1/2}.
\end{equation*}
For the occupation parameters, the direct-cutoff bound allows
\begin{equation*}
    q=1,\quad d=O\left(T^{\kappa+1}+\log_2(1/\veps)\right).
\end{equation*}
Alternatively, fixing any integer $d\geq2$, the diluted-cutoff bound
allows
\begin{equation*}
    q=O\left(T^{(\kappa+1)d/(d-1)}\veps^{-1/(d-1)}\right).
\end{equation*}
For the number of evolution steps, the coupling variation envelope and
the vacuum-initialized second-order simulation bound allow the
$q$-independent sufficient scaling
\begin{equation*}
    M_{\trot}=O\left(
        T^{5/2}\veps^{-1/2}\Lambda^2\right),
    \qquad \Lambda:=1+\sqrt{dN}\,T^c.
\end{equation*}
This estimate is conservative in $T$ and $d$; the explicit coefficient
in \eqref{eq:digital-error-constant} retains the decreasing inner
splitting error. Since $N,d$ are polynomially bounded, so are $\Lambda$
and the number of evolution steps. The detailed derivations are given
in \cref{app:resource-estimates}.

\paragraph{Resource scaling and tradeoff.}
Under these bath assumptions, the constructions therefore have
polynomial circuit width and depth in $T$ and $1/\veps$, provided the
elementary unitary implementations have polynomial cost:
\begin{equation*}
    W_{\mathrm{circ}},\;D_{\mathrm{circ}}
    =O\!\left(\operatorname{poly}(T,1/\veps)\right).
\end{equation*}
The direct construction uses fewer registers with larger local
dimension, while dilution reduces the required local dimension but
introduces additional bath registers and interactions. It therefore
offers implementation flexibility, rather than a guaranteed reduction in
Trotter error or circuit depth. The choice between them depends on the
cost of encoding and evolving these bath registers.

\section{Discussion}
\label{sec:discussion}

We first compare our construction with a related method called
ancilla-train algorithm, focusing on how the two methods represent
environmental memory. We then summarize the main results and discuss
directions for future work.

\subsection{Comparison with ancilla-train algorithm}

A closely related construction is the ancilla-train algorithm
\cite{christensen2025ancilla}, which considers the stationary two-time
correlation function
\begin{equation*}
    J(t-\tprime) = \tr\bigl[\rho_B B(t)B(\tprime)\bigr].
\end{equation*}
Both approaches share the same physical starting point: for a Gaussian
environment, the reduced dynamics is determined by the two-time
correlation kernel.  Both methods therefore replace the original
environment with an effective bath that reproduces this kernel.  The
essential difference lies in how the kernel is represented and
compressed into a finite auxiliary bath.

In the present work, we use a spectral decomposition of the kernel,
which yields the approximate bath
\begin{equation*}
    J(t-\tprime)
    =
    \sum_{n=1}^{\infty}\lambda_nu_n(t)u_n^*(\tprime),
    \qquad
    A_N(t)
    =
    \sum_{n=1}^{N}
    \left[
        \sqrt{\lambda_n}u_n(t)a_n
        +
        \sqrt{\lambda_n}u_n^*(t)a_n^\dagger
    \right].
\end{equation*}
By contrast, the ancilla-train algorithm uses a convolution
factorization and replaces the integral with a uniform-grid quadrature
of spacing $\Delta\xi$:
\begin{equation*}
    J(t-\tprime)
    =
    \int_{-\infty}^{\infty}
    g(t-\xi)g(\xi-\tprime)\,\rd\xi
    \approx
    \Delta\xi
    \sum_{n=-\infty}^{\infty}
    g(t-\xi_n)g(\xi_n-\tprime),
    \qquad
    \xi_n=n\Delta\xi,
\end{equation*}
where $g(t)$ is the jump correlator defined in
\cite{christensen2025ancilla}.  After applying a low-pass filter to
obtain $g_{\rm f}$ and restricting each bath mode to a memory window of
half-width $\tau_{\rm c}$, the finite train field is
\begin{equation*}
    A_{\rm T}(t)
    =
    \sqrt{\Delta\xi}
    \sum_{|t-\xi_n|\leq\tau_{\rm c}}
    \left[
        g_{\rm f}(t-\xi_n)a_n
        +
        g_{\rm f}^*(t-\xi_n)a_n^{\dagger}
    \right].
\end{equation*}

The two methods thus exploit different properties of environmental
memory. The ancilla train benefits from a well-localized jump
correlator, whereas our construction benefits from rapid decay of the
finite-interval kernel spectrum. The spectral construction also applies
directly to nonstationary kernels, for which translated copies of a
single jump correlator may not provide a suitable representation. The
practical choice depends on whether the bath correlations can be
represented more economically by a short memory window or by a small
number of collective modes. The combination of the two methods is also
an interesting future direction.

\subsection{Conclusion and outlook}

We developed a quantum algorithm for non-Markovian dynamics in Gaussian
environments with linear coupling and zero mean. Since the two-time
correlation kernel determines the reduced dynamics, we replace the
original bath with an effective bath built from the dominant spectral
modes of this kernel. An occupation cutoff makes each mode finite
dimensional, while coupling dilution allows smaller local dimensions at
the cost of additional bath registers. Combining this construction with
a time-dependent product formula gives a controlled total simulation
error. The resulting circuit width and depth are polynomial in the
simulation time and inverse target accuracy under the stated
spectral-decay and coupling-regularity assumptions, and provided
elementary unitaries admit polynomial-cost implementations.

The practical efficiency of the method depends on the spectrum of the
bath kernel, the accumulated couplings of the effective bath modes, and
their time variation. Rapid spectral decay favors a compact bath, while
the total circuit cost also depends on the occupation requirements and
implementation of its evolution. Important directions for future work
include numerical studies of representative physical kernels, adaptive
mode-dependent choices of the occupation cutoff and dilution factor,
extensions to multiple system--bath coupling operators, and
hardware-aware implementations of the resulting Hamiltonians.

\clearpage

\appendix

\section{Gaussian-Bath Equivalence} \label{app:kernel-equivalence}

Under the assumptions in \cref{sec:gaussian-model}, the initial bath
state $\rho_B$ is Gaussian and the bath coupling operator $B(t)$ is
linear in bosonic creation and annihilation operators with zero mean.
Wick's theorem gives vanishing odd moments and, for even moments,
\begin{equation*}
    \tr\bigl[\rho_B B(t_1)\cdots B(t_{2m})\bigr]
    =\sum_{\pi\in\frakP_2([2m])}
    \prod_{(i,j)\in\pi}G(t_i,t_j),
    \qquad G(t,s)=\tr[\rho_B B(t)B(s)].
\end{equation*}
Here $\frakP_2([2m])$ denotes the set of all disjoint pair partitions of
$\{1,\ldots,2m\}$, with $i<j$ in each pair $(i,j)$.

Thus the two-time correlation function $G(t,s)$ fixes all bath moments.
This gives the following Gaussian-bath equivalence principle
\cite{tamascelli2018nonperturbative}.

\begin{theorem}[Kernel equivalence]
    \label{thm:gaussian-kernel-realization}
    Let $(A(\cdot),\rho_A)$ and $(B(\cdot),\rho_B)$ be two Gaussian
    baths: each bath state is Gaussian and each coupling operator is
    Hermitian, linear in bosonic creation and annihilation operators,
    and has zero mean. Suppose
    \begin{equation*}
        \tr[\rho_A A(t)A(s)]
        =\tr[\rho_B B(t)B(s)]=G(t,s),
        \qquad 0\leq t,s\leq T.
    \end{equation*}
    For $X\in\{A,B\}$, let the joint evolution be generated by
    $H_X(t)=H_s\otimes I_X+S\otimes X(t)$, with common $H_s$, $S$, and
    initial system state $\rho_s(0)$, and with initial joint state
    $\rho_s(0)\otimes\rho_X$. Then the Schr\"odinger-picture reduced
    system states agree:
    \begin{equation*}
        \rho_s^{(A)}(t)=\rho_s^{(B)}(t),
        \qquad 0\leq t\leq T.
    \end{equation*}
\end{theorem}

\begin{proof}
    For $X\in\{A,B\}$, let $V_X(t)$ be the propagator generated by
    $H_X(t)$ with $V_X(0)=I$. To remove the free system evolution from
    the bath-only interaction picture, set
    $U_X(t)=(\re^{\ri H_s t}\otimes I_X)V_X(t)$. This propagator is
    generated by
    \begin{equation*}
        H_{I,X}(t)=S(t)\otimes X(t),
        \qquad S(t)=\re^{\ri H_s t}S\re^{-\ri H_s t}.
    \end{equation*}
    Expanding $U_X$ and $U_X^{\dagger}$ in Dyson series yields
    \begin{align*}
        \rho_{s,I}^{(X)}(t)
        &=\tr_X\!\left[U_X(t)(\rho_s(0)\otimes\rho_X)U_X(t)^\dagger\right] \\
        &=\sum_{m,n\geq0}(-\ri)^m\ri^n
        \int_{\Delta_m(t)}\!\rd^m\boldsymbol{t}
        \int_{\Delta_n(t)}\!\rd^n\boldsymbol{s}\;
        S(t_1)\cdots S(t_m)\rho_s(0)S(s_n)\cdots S(s_1) \\
        &\qquad\qquad\times
        \tr_X\!\left[\rho_X X(s_n)\cdots X(s_1)X(t_1)\cdots X(t_m)\right],
    \end{align*}
    where $\Delta_m(t)=\{t\geq t_1\geq\cdots\geq t_m\geq0\}$, and the
    zero-order integrals are understood as $1$. The bath trace vanishes
    when $m+n$ is odd; otherwise Wick's theorem expresses it as a sum
    of products of two-time correlation functions. Since both baths have
    the same kernel $G(t,s)$, their bath factors agree, as do the system
    factors and integration domains. Hence the double Dyson expansions
    give $\rho_{s,I}^{(A)}(t)=\rho_{s,I}^{(B)}(t)$. Finally,
    \begin{equation*}
        \rho_s^{(X)}(t)
        =\re^{-\ri H_s t}\rho_{s,I}^{(X)}(t)\re^{\ri H_s t},
    \end{equation*}
    so the same equality holds for the Schr\"odinger-picture reduced
    states.
\end{proof}

\section{Notation and Duhamel Expansions} \label{sec:preliminary}

We introduce the superoperator notation and norms used in the error
bounds, then state Duhamel's identities and their expansion form.
Combining these with a telescoping identity gives an expansion for
products of propagators separated by fixed superoperators.

\paragraph{Superoperator notation and norms.}

For the rank truncation error and bosonic occupation-cutoff error, we
use superoperators to write the evolution of the joint system-bath state
compactly. For a time-dependent Hamiltonian $H(t)$, the von Neumann
equation reads
\begin{equation*}
    \frac{\rd}{\rd t}\rho(t)=\calL(t)[\rho(t)],
    \qquad \calL(t)[\cdot]:=-\ri[H(t),\cdot],
\end{equation*}
where $\calL(t)$ is the Liouvillian. Its propagator is
\begin{equation*}
    \calU(t,s):=\calT\exp\left(\int_s^t\calL(\tau)\,\rd\tau\right),
    \qquad \calU(t):=\calU(t,0),
\end{equation*}
where $\calT$ denotes time ordering. Thus
\begin{equation*}
    \rho(t)=\calU(t)[\rho(0)]
    =U(t,0)\rho(0)U(t,0)^\dagger,
\end{equation*}
with $U(t,s)=\calT\exp\left(-\ri\int_s^t H(\tau)\,\rd\tau\right)$ the
joint unitary propagator. 

We write $\tracenorm{X}:=\tr\sqrt{X^\dagger X}$ for the trace norm and
define the induced norm of a superoperator $\calA$ by
\begin{equation*}
    \superopnorm{\calA}
    :=\sup_{\tracenorm{X}\leq1}\tracenorm{\calA[X]}.
\end{equation*}
Unitary evolution preserves the trace norm, so
$\superopnorm{\calU(t,s)}=1$. We also use that the partial trace cannot
increase the trace norm of a Hermitian operator.

\paragraph{Duhamel identities and expansion.}

Duhamel's identities express the difference between two propagators
in terms of the difference between their generators. We use both
equivalent forms below.

\begin{lemma}[Left and right Duhamel identities] \label{lem:duhamel}
    Let $H_1(t)$ and $H_2(t)$ be time-dependent Hamiltonians, with
    Liouvillians $\calL_1(t)$ and $\calL_2(t)$ and propagators
    $\calU_1(t,s)$ and $\calU_2(t,s)$, respectively. Set
    \begin{equation*}
        \Delta H(t):=H_1(t)-H_2(t),
        \qquad
        \Delta\calL(t):=\calL_1(t)-\calL_2(t)
        =-\ri[\Delta H(t),\cdot].
    \end{equation*}
    Then, for any $t\geq s$,
    \begin{align*}
        \calU_1(t,s)-\calU_2(t,s)
        &=
        \int_s^t
        \calU_1(t,\tau)\Delta\calL(\tau)\calU_2(\tau,s)\,\rd\tau
        \qquad \text{(left Duhamel identity)} \\
        &=
        \int_s^t
        \calU_2(t,\tau)\Delta\calL(\tau)\calU_1(\tau,s)\,\rd\tau
        \qquad \text{(right Duhamel identity)} .
    \end{align*}
\end{lemma}

\begin{proof}
    Differentiating the product of the two propagators gives
    \begin{equation*}
        \frac{\rd}{\rd\tau}
        \left[\calU_1(t,\tau)\calU_2(\tau,s)\right]
        =
        -\calU_1(t,\tau)\Delta\calL(\tau)\calU_2(\tau,s).
    \end{equation*}
    Integrating from $s$ to $t$ yields the first identity. Interchanging
    $1$ and $2$ and using $\calL_2-\calL_1=-\Delta\calL$ gives the
    second.
\end{proof}

The same argument gives the corresponding identities for the unitary
propagators $U_j(t,s)$ generated by $H_j(t)$:
\begin{align*}
    U_1(t,s)-U_2(t,s)
    =-\ri\int_s^t
    U_1(t,\tau)\Delta H(\tau)U_2(\tau,s)\,\rd\tau 
    =-\ri\int_s^t
    U_2(t,\tau)\Delta H(\tau)U_1(\tau,s)\,\rd\tau.
\end{align*}

\begin{remark}[Duhamel expansion]
    For a splitting $H(t)=H_0(t)+R(t)$, let $\calU$ and $\calU_0$
    be the propagators generated by $H$ and $H_0$, respectively, and
    set $\calR(t)[X]:=-\ri[R(t),X]$. Then \cref{lem:duhamel} gives
    \begin{equation*}
        \calU(t,s)
        =
        \calU_0(t,s) +
        \int_s^t
        \calU(t,\tau)\calR(\tau)\calU_0(\tau,s)\,\rd\tau.
    \end{equation*}
    This replaces $\calU$ by $\calU_0$ plus an exact integral
    remainder containing one explicit factor of $\calR$. We call
    this a Duhamel expansion step.
\end{remark}

\paragraph{Telescoping Duhamel expansion.}

To extend a Duhamel expansion step to a product containing fixed
insertions, we first use the following algebraic identity. It expresses
the difference between two ordered products as a sum of terms, each
containing the difference of one pair of factors.

\begin{lemma}[Telescoping identity]
    \label{lem:telescoping}
    Let $X_0,\ldots,X_m$ and $Y_0,\ldots,Y_m$ be operators, or more
    generally superoperators, on the same vector space. Then
    \begin{equation*}
        X_mX_{m-1}\cdots X_0-Y_mY_{m-1}\cdots Y_0
        =
        \sum_{a=0}^{m}
        X_m\cdots X_{a+1}
        (X_a-Y_a)
        Y_{a-1}\cdots Y_0 ,
    \end{equation*}
    where empty products are the identity.
\end{lemma}

\begin{proof}
    Insert the intermediate products obtained by replacing the factors
    one at a time from right to left:
    \begin{align*}
        X_m\cdots X_0-Y_m\cdots Y_0
        =
        \sum_{a=0}^{m}
        \left(
        X_m\cdots X_{a+1}X_aY_{a-1}\cdots Y_0
        -
        X_m\cdots X_{a+1}Y_aY_{a-1}\cdots Y_0
        \right).
    \end{align*}
    The intermediate products cancel telescopically. Factoring
    $X_a-Y_a$ in each summand proves the claim.
\end{proof}

Combining \cref{lem:telescoping,lem:duhamel} gives the following
expansion. Each integral term exposes one generator difference between
fixed insertions. Propagators before that difference use the comparison
generators, while those after it retain the original generators.

\begin{lemma}[Telescoping Duhamel expansion]
    \label{lem:telescoping-duhamel}
    Fix times $t_0\leq t_1\leq\cdots\leq t_{m+1}$.
    For each $0\leq k\leq m$, let $H_k(\tau)$ and $\widehat H_k(\tau)$
    be time-dependent Hamiltonians on the same Hilbert space, with
    Liouvillians $\calL_k(\tau)$ and $\widehat{\calL}_k(\tau)$.
    Let $\calV_k$ and $\widehat{\calV}_k$ denote the respective
    evolution propagators on $[t_k,t_{k+1}]$. Write
    \begin{equation*}
        \Delta\calL_k(\tau):=\calL_k(\tau)-\widehat{\calL}_k(\tau),
        \qquad
        \calV_k:=\calV_k(t_{k+1},t_k),
        \qquad
        \widehat{\calV}_k:=\widehat{\calV}_k(t_{k+1},t_k).
    \end{equation*}
    Let $\calA_1,\ldots,\calA_m$ be fixed bounded superoperators
    acting on the same operator space. Then
    \begin{align*}
        &\quad \calV_m\calA_m\cdots\calV_1\calA_1\calV_0 
        =\widehat{\calV}_m\calA_m\cdots
            \widehat{\calV}_1\calA_1\widehat{\calV}_0\\
        &\quad+\sum_{k=0}^{m}\int_{t_k}^{t_{k+1}}
        \bigl(\calV_m\calA_m\cdots
            \calV_{k+1}\calA_{k+1}\bigr)
        \calV_k(t_{k+1},\tau)\Delta\calL_k(\tau) 
        \widehat{\calV}_k(\tau,t_k)
        \bigl(\calA_k\widehat{\calV}_{k-1}\cdots
            \calA_1\widehat{\calV}_0\bigr)\,\rd \tau .
    \end{align*}
    Empty products are the identity: the first parenthesized product
    is the identity for $k=m$, and the last for $k=0$.
    For $m=0$, the formula reduces to a single Duhamel expansion step.
\end{lemma}

\begin{proof}
    Apply \cref{lem:telescoping} with
    \begin{equation*}
        X_0=\calV_0,\qquad Y_0=\widehat{\calV}_0,
        \qquad
        X_k=\calV_k\calA_k,\qquad
        Y_k=\widehat{\calV}_k\calA_k\quad(1\leq k\leq m).
    \end{equation*}
    For $k\geq1$, the difference factor is
    $(\calV_k-\widehat{\calV}_k)\calA_k$; for $k=0$, it is
    $\calV_0-\widehat{\calV}_0$. In each case, substitute
    \begin{equation*}
        \calV_k-\widehat{\calV}_k
        =\int_{t_k}^{t_{k+1}}
        \calV_k(t_{k+1},\tau)\Delta\calL_k(\tau)
        \widehat{\calV}_k(\tau,t_k)\,\rd \tau,
    \end{equation*}
    which follows from \cref{lem:duhamel}. Keeping the surrounding
    factors in place and moving the comparison product to the right
    gives the claimed expansion.
\end{proof}

\section{Rank truncation error}
\label{app:rank-truncation-error}

This appendix bounds the reduced state error caused by retaining only
the leading $N$ spectral modes of the auxiliary Gaussian bath.
The bound depends on the discarded eigenvalues and the system coupling
strength, and provides the rank-selection criterion used in
\eqref{eq:rank-truncation-choice}.

We first state the Gaussian correlation perturbation bound in
physical-time notation, allowing the time-dependent Gaussian bath
couplings used in our auxiliary construction.

\begin{theorem}[Gaussian correlation perturbation bound \cite{liu2025error}]
    \label{thm:gaussian-correlation-perturbation}
    Consider two centered Gaussian baths with Hermitian coupling
    operators linear in bosonic creation and annihilation operators,
    and bounded correlation kernels $G$ and $\widetilde G$ on
    $[0,T]^2$. Suppose that both joint evolutions start from product
    states with the same initial system state, have the same system
    Hamiltonian $H_s$ and bounded Hermitian system coupling $S$, and
    admit unitary propagators. Let $\rho_s^{(G)}(T)$ and
    $\rho_s^{(\widetilde G)}(T)$ denote their reduced system states.
    Then, for every bounded Hermitian system observable $O$,
    \begin{equation*}
        \left|\tr\!\left[
            O\bigl(\rho_s^{(G)}(T)-\rho_s^{(\widetilde G)}(T)\bigr)
        \right]\right|
        \leq \opnorm{O}\left[
            \exp\!\left(
                2\opnorm{S}^2
                \int_0^T\int_0^T|G(t,s)-\widetilde G(t,s)|
                \,\rd t\,\rd s
            \right)-1
        \right].
    \end{equation*}
\end{theorem}

\begin{theorem}[Rank truncation of the bath]
    \label{thm:rank-truncation-error}
    Let $G$ be a bounded positive Hermitian bath kernel on $[0,T]$ with
    spectral decomposition
    \begin{equation*}
        G(t,s)=\sum_{n=1}^{\infty}\lambda_n u_n(t)u_n^*(s),
        \qquad
        \lambda_1\geq\lambda_2\geq\cdots\geq0,
    \end{equation*}
    and let $G_N$ be its rank-$N$ truncation. Let $\rho_s(T)$ and
    $\rho_s^{(N)}(T)$ be the reduced system states generated from the
    same initial system state by the Gaussian baths with kernels $G$ and
    $G_N$, respectively, under the assumptions of
    \cref{thm:gaussian-correlation-perturbation}. Assume that the
    spectral tail bound is finite:
    \begin{equation*}
        \eta_N
        :=
        T\left(\sum_{n>N}\lambda_n^2\right)^{1/2}<\infty.
    \end{equation*}
    Then
    \begin{equation*}
        \tracenorm{\rho_s(T)-\rho_s^{(N)}(T)}
        \leq
        \exp\!\left(2\opnorm{S}^2\eta_N\right)-1.
    \end{equation*}
\end{theorem}

\begin{proof}
    Set $\Delta_N=G-G_N$ and
    $\Lonenorm{\Delta_N}:=\int_0^T\int_0^T
    |\Delta_N(t,s)|\,\rd t\,\rd s$.
    Applying \cref{thm:gaussian-correlation-perturbation} to $G$ and
    $G_N$ and using trace-norm duality for Hermitian operators,
    \begin{equation*}
        \tracenorm{X}
        =\sup_{O=O^\dagger,\,\opnorm{O}\leq1}|\tr(OX)|,
    \end{equation*}
    gives
    \begin{equation*}
        \tracenorm{\rho_s(T)-\rho_s^{(N)}(T)}
        \leq\exp\!\left(2\opnorm{S}^2\Lonenorm{\Delta_N}\right)-1.
    \end{equation*}
    Finally, Cauchy--Schwarz on $[0,T]^2$ and orthonormality of the
    kernels $u_n(t)u_n^*(s)$ imply
    \begin{equation*}
        \Lonenorm{\Delta_N}
        \leq T\left(
            \int_0^T\int_0^T|\Delta_N(t,s)|^2\,\rd t\,\rd s
        \right)^{1/2}
        =T\left(\sum_{n>N}\lambda_n^2\right)^{1/2}
        =\eta_N.
    \end{equation*}
    Substitution proves the claimed bound.
\end{proof}

\section{Occupation-Cutoff and Coupling-Dilution Errors}
\label{app:occupation-cutoff-error}

This appendix gives two bounds on the error introduced by truncating
local bosonic occupations. We treat the diluted construction with
$q$ copies per retained mode; the direct cutoff corresponds to $q=1$.
The two bounds control the same error but emphasize different ways to
reduce it: increasing the local dimension $d$ or increasing the number
of copies $q$.

\subsection{Error bounds}

\begin{theorem}[Bosonic occupation-cutoff error]
    \label{thm:diluted-d-level-error}
    Let $S$ be bounded and let the retained coupling functions $g_n$
    be integrable on $[0,T]$. Let $\rho_s^{(N)}(t)$ be the reduced
    system state generated by $H_s\otimes I+S\otimes A_N(t)$, with
    initial state $\rho_s(0)\otimes\ket{\bzero}\bra{\bzero}$.
    For integers $q\geq1$ and $d\geq2$, let
    $\rho_s^{(N,q,d)}(t)$ be the reduced system state generated by
    $H_s\otimes I+S\otimes A_{N,q,d}(t)$, with the same initial
    system state and all $Nq$ finite-level modes initially in vacuum.
    Define the accumulated coupling
    \begin{equation*}
        \beta_{n,t}:=\opnorm{S}\int_0^t|g_n(u)|\,\rd u.
    \end{equation*}
    Then, for every $t\in[0,T]$, both bounds hold:
    \begin{align}
        \tracenorm{\rho_s^{(N)}(t)-\rho_s^{(N,q,d)}(t)}
        &\leq
        \frac{2^d}{q^{d/2-1}\sqrt{d!}}
        \sum_{n=1}^N\beta_{n,t}^{d},
        \label{eq:occupation-factorial-bound}\\
        \tracenorm{\rho_s^{(N)}(t)-\rho_s^{(N,q,d)}(t)}
        &\leq
        \frac{2^d}{q^{d-1}}
        \sum_{n=1}^N\beta_{n,t}^{2d}.
        \label{eq:occupation-dilution-bound}
    \end{align}
\end{theorem}

For fixed $N$, $q$, and accumulated couplings, the factorial in
\eqref{eq:occupation-factorial-bound} ensures convergence as
$d\to\infty$. For fixed $d\geq2$,
\eqref{eq:occupation-dilution-bound} instead gives convergence as
$q\to\infty$, with error bounded by $O(q^{1-d})$. In particular,
$d=2$ gives the bound $4\sum_n\beta_{n,t}^4/q$.
Either estimate may be used, so the smaller bound can be chosen for
given $d$ and $q$. Setting $q=1$ in the first bound gives the direct
cutoff estimate.

\begin{corollary}[Error of the direct $d$-level bath]
    \label{cor:direct-d-level-error}
    Under the assumptions of \cref{thm:diluted-d-level-error}, let
    $\rho_s^{(N,d)}(t)$ be the reduced system state generated by the
    direct $d$-level bath $A_{N,d}(t)$. Then, for every $t\in[0,T]$,
    \begin{equation*}
        \tracenorm{\rho_s^{(N)}(t)-\rho_s^{(N,d)}(t)}
        \leq\frac{2^d}{\sqrt{d!}}
        \sum_{n=1}^N\beta_{n,t}^{d}.
    \end{equation*}
\end{corollary}

\begin{proof}
    Set $q=1$ in \eqref{eq:occupation-factorial-bound} and use
    $A_{N,1,d}=A_{N,d}$.
\end{proof}

For these error bounds, we first express the occupation-cutoff error
through transitions at the cutoff boundary. Then we use Duhamel
expansion to extract the scaling with the dilution factor. The
first-crossing expansion exposes one insertion of each local transition
Liouvillian, yielding \eqref{eq:occupation-factorial-bound}. Using
cancellation under the bath trace, the second-crossing expansion exposes
two insertions of each transition and yields
\eqref{eq:occupation-dilution-bound}. 

\subsection{Error decomposition}

We first express the occupation-cutoff error through transitions at the
cutoff boundary and establish the local norm bounds used in the 
error estimates.

\paragraph{Exact and truncated evolutions.}

The exact bosonic diluted bath has the same correlation kernel as the
retained $N$-mode bosonic bath and therefore generates the same reduced
state $\rho_s^{(N)}(t)$. We can thus obtain the occupation-cutoff error
by comparing the exact diluted evolution with its $d$-level truncation.
Both evolutions start from
\begin{equation*}
    \rho_{sb}(0):=\rho_s(0)\otimes\ket{\bzero}\bra{\bzero},
    \qquad
    \ket{\bzero}:=\bigotimes_{n=1}^{N}\bigotimes_{r=1}^{q}\ket{0}_{n,r}.
\end{equation*}
Let $H_{\dil}(t)$ and $H_{\dil,d}(t)$ denote the exact and truncated
diluted Hamiltonians, respectively:
\begin{equation*}
    H_{\dil}(t):=H_s\otimes I+S\otimes A_{N,q}(t),
    \qquad
    H_{\dil,d}(t):=H_s\otimes I+S\otimes A_{N,q,d}(t).
\end{equation*}
Let $\calU_{\dil}(t)$ and $\calU_{\dil,d}(t)$ be the corresponding
superoperator propagators. For any trace-class system operator $X$,
define the difference of the reduced evolution channels,
\begin{equation*}
    \calE_{\dil,d}(t)[X]
    :=\tr_b\bigl[
        \left(\calU_{\dil}(t)-\calU_{\dil,d}(t)\right)
        \left[X\otimes\ket{\bzero}\bra{\bzero}\right]
    \bigr].
\end{equation*}
Then the reduced-state error is
\begin{equation*}
    \tracenorm{\rho_s^{(N)}(t)-\rho_s^{(N,q,d)}(t)}
    =\tracenorm{
        \tr_b\bigl[
            \left(\calU_{\dil}(t)-\calU_{\dil,d}(t)\right)
            [\rho_{sb}(0)]
        \bigr]
    }
    =\tracenorm{\calE_{\dil,d}(t)[\rho_s(0)]}.
\end{equation*}

\paragraph{Omitted transitions.}

To express these Hamiltonians in terms of adjacent local transitions,
we extend the transition Hamiltonians $H_{n,r,j}(t)$ defined in
\eqref{eq:finite-bath-transition} to all occupation edges $j\geq1$:

\begin{equation*}
    H_{n,r,j}(t)
    :=
    S\otimes
    \frac{\sqrt j}{\sqrt q}
    \left(
        g_n(t)(\ket{j-1}\bra j)_{n,r}
        +
        g_n^*(t)(\ket j\bra{j-1})_{n,r}
    \right),
    \qquad j\geq1 .
\end{equation*}
The identity on all bath modes other than $(n,r)$ is implicit. The two
Hamiltonians differ only by the missing copied-mode transitions at and
above the cutoff boundary:
\begin{equation*}
    H_{\dil}(t) - H_{\dil,d}(t)
    =
    \sum_{n=1}^{N}\sum_{r=1}^{q}\sum_{j=d}^{\infty} H_{n,r,j}(t).
\end{equation*}
Define the associated transition superoperator by
\begin{equation*}
    \calL_{n,r,j}(t)[X]:=-\ri[H_{n,r,j}(t),X].
\end{equation*}
Let
\begin{equation*}
    \calL_{\dil}(t)[\cdot] := -\ri[H_{\dil}(t), \cdot],
    \quad
    \calL_{\dil,d}(t)[\cdot] := -\ri[H_{\dil,d}(t), \cdot],
\end{equation*}
then the Liouvillian difference is
\begin{equation*}
    \calL_{\dil}(t) - \calL_{\dil,d}(t) 
    = \sum_{n=1}^{N}\sum_{r=1}^{q}\sum_{j=d}^{\infty} \calL_{n,r,j}(t).
\end{equation*}

Removing $H_{n,r,j}$ from the exact diluted Hamiltonian prevents
transitions between occupations below $j$ and those at or above $j$
in mode $(n,r)$. We use this property in the error analysis below.

\paragraph{Boundary reduction.}

We now decompose the channel error into copied-mode contributions. Since
the $d$-level evolution preserves the local subspace of each copied mode
initialized in vacuum, only the boundary transition
$\ket{d-1}\leftrightarrow\ket d$ can contribute.

\begin{lemma} \label{lem:diluted-boundary-reduction}
    The reduced-channel difference admits the decomposition
    \begin{equation*}
        \calE_{\dil, d}(t)
        =
        \sum_{n=1}^N\sum_{r=1}^q
        \calE_{n,r}^{(d)}(t),
    \end{equation*}
    where
    \begin{equation*}
        \calE_{n,r}^{(d)}(t)[\rho_s(0)]
        :=
        \int_0^t \rd \tau\,
        \tr_b\left[
            \calU_{\dil}(t,\tau)
            \calL_{n,r,d}(\tau)
            \calU_{\dil,d}(\tau)
            [\rho_{sb}(0)]
        \right].
    \end{equation*}
\end{lemma}

\begin{proof}
    By the Duhamel identity in \cref{lem:duhamel},
    \begin{align*}
        \calU_{\dil}(t)-\calU_{\dil,d}(t)
        &=
        \int_0^t
        \calU_{\dil}(t,\tau)
        \left(\calL_{\dil}(\tau)-\calL_{\dil,d}(\tau)\right)
        \calU_{\dil,d}(\tau)\,\rd\tau \\
        &=
        \sum_{n=1}^N\sum_{r=1}^q
        \sum_{j=d}^{\infty}
        \int_0^t
        \calU_{\dil}(t,\tau)
        \calL_{n,r,j}(\tau)
        \calU_{\dil,d}(\tau)\,\rd\tau .
    \end{align*}
    The $d$-level evolution $\calU_{\dil,d}$ preserves the local
    subspace $\operatorname{span}\{\ket{0},\ldots,\ket{d-1}\}_{n,r}$
    when the copied mode starts in $\ket{0}$. Then we have
    $\calL_{n,r,j}(\tau)\calU_{\dil,d}(\tau)[\rho_{sb}(0)]=0$ for every
    $j\geq d+1$, since $H_{n,r,j}$ acts only between the levels $j-1$
    and $j$, both outside this local subspace. Hence only the boundary
    transition $j=d$ contributes. Applying the identity to
    $\rho_{sb}(0)$ and tracing out the bath gives the claimed
    decomposition.
\end{proof}

\paragraph{Local transition bounds.}

The following bounds control the norm of each transition Hamiltonian
and its commutator superoperator.

\begin{lemma}[Local ladder Liouvillian bounds]
    \label{lem:local-ladder-liouvillian-bounds}
    For each copied mode $(n,r)$ and $j\geq1$, the local transition
    operators satisfy
    \begin{equation*}
        \opnorm{H_{n,r,j}(t)}
        \leq
        \frac{\sqrt j}{\sqrt q}\,\opnorm{S}\,|g_n(t)|,
        \qquad
        \superopnorm{\calL_{n,r,j}(t)}
        \leq
        \frac{2\sqrt j}{\sqrt q}\,\opnorm{S}\,|g_n(t)|.
    \end{equation*}
    The superoperator norm is taken on trace-class operators of the
    full system-bath space.
\end{lemma}

\begin{proof}
    On the two-level subspace
    $\operatorname{span}\{\ket{j-1},\ket{j}\}_{n,r}$, the local bath
    factor has norm $\sqrt{j/q}|g_n(t)|$.  Tensoring with $S$ and the
    identities on the remaining modes gives the first bound. The second
    follows from
    \begin{equation*}
        \tracenorm{\calL_{n,r,j}(t)[X]}
        =\tracenorm{[H_{n,r,j}(t),X]}
        \leq 2\opnorm{H_{n,r,j}(t)}\tracenorm{X}
    \end{equation*}
    for every trace-class operator $X$.
\end{proof}

We next use these bounds to estimate each boundary contribution.

\subsection{Proof of the factorial bound}

For each copied mode $(n,r)$, the boundary-reduction formula exposes
one local Liouvillian $\calL_{n,r,d}$, whose norm bound contains a
factor $q^{-1/2}$. Bounding the remaining evolution directly would
retain only this one factor. Starting from vacuum, however, reaching
the boundary also requires crossing transitions $1,\ldots,d-1$.
We therefore apply Duhamel's identity repeatedly to the evolution
before the boundary transition, exposing one insertion of each lower
transition. Each contributes another factor $q^{-1/2}$, giving
$q^{-d/2}$ per copied mode and $q^{1-d/2}$ after summing over the $q$
copies. 

Next, we make such an observation into a rigorous expansion in the
following lemma. The expansion is exact, with further interactions
retained in the intervening propagators. Its proof is given in
\cref{sec:crossing-expansion}.

\begin{lemma}[First-crossing expansion]
    \label{lem:first-crossing-expansion}
    For each fixed copied mode $(n,r)$, the copied-mode error
    contribution admits the expansion
    \begin{align*}
        \calE_{n,r}^{(d)}(t)[\rho_s(0)]
        &=
        \int_{0\leq t_1\leq\cdots\leq t_d\leq t}
        \tr_b\Bigl[
            \calV_d(t,t_d)
            \calL_{n,r,d}(t_d) \times
            \calV_{d-1}(t_d,t_{d-1})
            \calL_{n,r,d-1}(t_{d-1}) \times
            \cdots
        \\
        &\qquad\qquad\qquad\qquad\qquad
            \cdots
            \calV_1(t_2,t_1)
            \calL_{n,r,1}(t_1) \times 
            \calV_0(t_1,0)[\rho_{sb}(0)]
        \Bigr]
        \,\rd t_1\cdots\rd t_d .
    \end{align*}
    Here $\calV_d=\calU_{\dil}$, while for $0\leq a\leq d-1$,
    $\calV_a$ is the Liouvillian propagator generated by
    $H_{\dil,d}$ with the copied-mode transitions
    $H_{n,r,a+1},\ldots,H_{n,r,d-1}$ removed, i.e.,
    \begin{equation*}
        H_{\dil,d}(t) - \sum_{j=a+1}^{d-1} H_{n,r,j}(t).
    \end{equation*}
    All other copied modes are unchanged. For $a=d-1$, the sum is
    empty, so $\calV_{d-1}=\calU_{\dil,d}$.
\end{lemma}

\begin{proof}[Proof of \cref{eq:occupation-factorial-bound}]
    The propagators in \cref{lem:first-crossing-expansion}
    preserve trace norm and Hermiticity. Each $\calL_{n,r,j}$ also
    preserves Hermiticity, so the final partial trace is contractive
    on the resulting operators. Using
    \cref{lem:local-ladder-liouvillian-bounds}, we obtain
    \begin{align*}
        \tracenorm{\calE_{n,r}^{(d)}(t)[\rho_s(0)]}
        &\leq
        \int_{0\leq t_1\leq\cdots\leq t_d\leq t}
        \prod_{j=1}^{d}
        \frac{2\sqrt j\,\opnorm{S}\,|g_n(t_j)|}{\sqrt q}
        \,\rd t_1\cdots\rd t_d
        \\
        &=
        \frac{2^d}{q^{d/2}\sqrt{d!}}
        \left(
            \opnorm{S}\int_0^t |g_n(u)|\,\rd u
        \right)^d .
    \end{align*}
    Here the transition factors give
    $\prod_{j=1}^d\sqrt j=\sqrt{d!}$, while symmetry of
    $\prod_{j=1}^d|g_n(t_j)|$ makes its integral over the ordered
    simplex equal to $1/d!$ times its integral over $[0,t]^d$.
    Together these yield the factor $1/\sqrt{d!}$.
    Summing over $r=1,\ldots,q$ and $n=1,\ldots,N$ gives
    \begin{align*}
        \tracenorm{\rho_s^{(N)}(t)-\rho_s^{(N,q,d)}(t)}
        &\leq
        \frac{2^d}{q^{d/2-1}\sqrt{d!}}
        \sum_{n=1}^{N}
        \left(\opnorm{S} \int_0^t |g_n(u)|\,\rd u\right)^d .
    \end{align*}
    This proves \eqref{eq:occupation-factorial-bound}.
\end{proof}

\subsection{Proof of the dilution bound}

Starting from the first-crossing expansion, we expose second occurrences
of the transition labels in descending order $d,d-1,\ldots,1$. At each
step, the first insertion of Liouvillian acts on an operator supported
below the corresponding cut and produces an off-diagonal operator.
Removing that transition from all later propagators preserves this
sector, so the bath partial trace vanishes. The telescoping Duhamel
expansion therefore leaves terms with a second explicit insertion, while
further interactions remain in the intervening propagators.

\begin{definition}[Admissible word for Liouvillians]
    A sequence $\omega=(\omega_1,\ldots,\omega_{2d})$ is called
    admissible if each label $1,\ldots,d$ occurs exactly twice and
    the first occurrences appear in the order $1,\ldots,d$.
    Here $\omega_a$ labels the explicit Liouvillian insertion at
    time $t_a$. We denote the set of these words by $\Omega_d$.
\end{definition}

For $d=2$, the admissible words are
\begin{equation*}
    (1,1,2,2),\qquad (1,2,1,2),\qquad (1,2,2,1).
\end{equation*}
These words record the chronological order of explicit Liouvillian
factors; further interactions remain in the propagators. Starting from
$(1,\ldots,d)$, insert second occurrences in descending label order. The
second $j$ has $2(d-j)+1$ possible placements after its first
occurrence. Every admissible word is obtained uniquely: deleting second
occurrences in increasing label order reverses the construction. Hence
the number of words in $\Omega_d$ is
\begin{equation*}
    |\Omega_d|
    =\prod_{j=1}^{d}\bigl(2(d-j)+1\bigr)
    =(2d-1)!!.
\end{equation*}

\begin{lemma}[Second-crossing expansion]
    \label{lem:second-crossing-expansion}
    For each copied mode $(n,r)$, the error contribution admits the
    expansion
    \begin{align*}
        \calE_{n,r}^{(d)}(t)[\rho_s(0)]
        &=\sum_{\omega\in\Omega_d}
        \int_{0\leq t_1\leq\cdots\leq t_{2d}\leq t}
        \tr_b\Bigl[
            \calV_{\omega,2d}(t,t_{2d})
            \calL_{n,r,\omega_{2d}}(t_{2d})\\
        &\qquad\qquad\times
            \calV_{\omega,2d-1}(t_{2d},t_{2d-1})
            \calL_{n,r,\omega_{2d-1}}(t_{2d-1})
            \cdots \\
        &\qquad\qquad\times
            \calV_{\omega,1}(t_2,t_1)
            \calL_{n,r,\omega_1}(t_1) \times
            \calV_{\omega,0}(t_1,0)[\rho_{sb}(0)]
        \Bigr]
        \,\rd t_1\cdots\rd t_{2d}.
    \end{align*}
    Here $\calV_{\omega,a}$, $a=0,\ldots,2d$, are unitary superoperator
    propagators. Each term in the summation contains exactly two
    explicit insertions of every local transition Liouvillian
    $\calL_{n,r,j}$, $j=1,\ldots,d$ and there are $|\Omega_d|=(2d-1)!!$
    terms.
\end{lemma}

The proof of the expansion is deferred to
\cref{sec:crossing-expansion}. We now use it to prove
\eqref{eq:occupation-dilution-bound}.

\begin{proof}[Proof of \cref{eq:occupation-dilution-bound}]
    Fix $(n,r)$. Each term in \cref{lem:second-crossing-expansion}
    contains two explicit insertions of every local transition
    Liouvillian. Propagators there preserve trace norm and Hermiticity,
    the transition superoperators preserve Hermiticity, and the final
    partial trace is contractive. Applying
    \cref{lem:local-ladder-liouvillian-bounds} and summing over at most
    $(2d-1)!!$ terms gives
    \begin{align*}
        \tracenorm{\calE_{n,r}^{(d)}(t)[\rho_s(0)]}
        &\leq
        (2d-1)!!
        \int_{0\leq t_1\leq\cdots\leq t_{2d}\leq t}
        \prod_{j=1}^{d}
        \left(
            \frac{2\sqrt{j}\,\opnorm{S}}{\sqrt q}
        \right)^2
        \prod_{a=1}^{2d}|g_n(t_a)|
        \,\rd t_1\cdots\rd t_{2d} \\
        &=
        (2d-1)!!
        \frac{4^d d!\,\opnorm{S}^{2d}}{q^d}
        \int_{0\leq t_1\leq\cdots\leq t_{2d}\leq t}
        \prod_{a=1}^{2d}|g_n(t_a)|
        \,\rd t_1\cdots\rd t_{2d}.
    \end{align*}
    The integrand is symmetric, so the ordered-simplex integral is
    \begin{equation*}
        \int_{0\leq t_1\leq\cdots\leq t_{2d}\leq t}
        \prod_{a=1}^{2d}|g_n(t_a)|
        \,\rd t_1\cdots\rd t_{2d}
        =
        \frac{1}{(2d)!}
        \left(\int_0^t |g_n(u)|\,\rd u\right)^{2d} .
    \end{equation*}
    Using $(2d-1)!!=\frac{(2d)!}{2^d d!}$ we obtain
    \begin{equation*}
        \tracenorm{\calE_{n,r}^{(d)}(t)[\rho_s(0)]}
        \leq
        \frac{2^d}{q^d}
        \left(\opnorm{S} \int_0^t |g_n(u)|\,\rd u\right)^{2d}.
    \end{equation*}
    Summing over the $q$ copies changes the factor $q^{-d}$ to
    $q^{1-d}$. Summing also over $n=1,\ldots,N$ gives
    \begin{align*}
        \tracenorm{\rho_s^{(N)}(t)-\rho_s^{(N,q,d)}(t)}
        \leq
        \sum_{n=1}^N\sum_{r=1}^q
        \tracenorm{\calE_{n,r}^{(d)}(t)[\rho_s(0)]}
        \leq
        \frac{2^d}{q^{d-1}}
        \sum_{n=1}^N
        \left(\opnorm{S} \int_0^t |g_n(u)|\,\rd u\right)^{2d}.
    \end{align*}
    This proves the \eqref{eq:occupation-dilution-bound}.
\end{proof}

\section{Proofs of the Crossing Expansions}
\label{sec:crossing-expansion}

We prove the first- and second-crossing expansion formulas used in
\cref{app:occupation-cutoff-error}. Throughout this appendix, fix one
copied mode $(n,r)$ and write
\begin{equation*}
    H_j:=H_{n,r,j},
    \qquad
    \calL_j:=\calL_{n,r,j},
    \qquad
    \rho_0:=\rho_{sb}(0).
\end{equation*}
The transition Hamiltonians and Liouvillians are those defined in
\cref{app:occupation-cutoff-error}.

\begin{definition}[Blocks associated with a cut]
    Fix an edge $j\geq1$ of the local occupation chain of the copied
    mode $(n,r)$. Define the lower and upper local projections
    \begin{equation*}
        P_{<j}:=\sum_{a=0}^{j-1}\ket{a}\bra{a},
        \qquad
        P_{\geq j}:=I-P_{<j}.
    \end{equation*}
    We use the same symbols after tensoring these projections with the
    identity on the system and on all other bath modes. The
    decomposition of the joint system-bath Hilbert space $\scrH$ as
    \begin{equation*}
        \scrH=P_{<j}\scrH \oplus P_{\geq j}\scrH
    \end{equation*}
    is called the cut at the edge $j-1\leftrightarrow j$. For an
    operator $X$, its diagonal and off-diagonal parts relative to this
    cut are
    \begin{align*}
        X_{\mathrm{diag}(j)}
        :=
        P_{<j}XP_{<j}+P_{\geq j}XP_{\geq j},
        \qquad
        X_{\mathrm{off}(j)}
        :=
        P_{<j}XP_{\geq j}+P_{\geq j}XP_{<j}.
    \end{align*}
    Together they give the four blocks associated with this
    decomposition of $\scrH$. Thus $X$ is below the cut if
    $X=P_{<j}XP_{<j}$, above the cut if $X=P_{\geq j}XP_{\geq j}$, and
    off-diagonal across the cut if $X=X_{\mathrm{off}(j)}$.
\end{definition}

\begin{lemma}[Cut rules] \label{lem:cut-invariance}
    Fix the cut at the edge $j-1\leftrightarrow j$.

    \begin{itemize}
        \item \emph{No-crossing rule.} Suppose a Hamiltonian $H(t)$
        is block-diagonal relative to the cut, that is,
        \begin{equation*}
            P_{<j}H(t)P_{\geq j}=0,
            \qquad
            P_{\geq j}H(t)P_{<j}=0
            \quad
            \text{for all }t .
        \end{equation*}
        Then its evolution preserves each of the four operator blocks.
        In particular, an operator supported below the cut remains
        below it, and an off-diagonal operator remains off-diagonal.

        \item \emph{Crossing rule.} The transition Liouvillian
        $\calL_j(t)[\cdot]=-\ri[H_j(t),\cdot]$ maps diagonal operators
        to off-diagonal operators relative to the cut. Thus, if
        $X=X_{\mathrm{diag}(j)}$, then
        \begin{equation*}
            \calL_j(t)[X]
            =
            P_{\geq j}\calL_j(t)[X]P_{<j}
            +
            P_{<j}\calL_j(t)[X]P_{\geq j}.
        \end{equation*}

        \item \emph{Trace cancellation.} Every trace-class operator
        that is off-diagonal across the cut has zero bath partial trace:
        \begin{equation*}
            X=X_{\mathrm{off}(j)}
            \quad\Longrightarrow\quad
            \tr_b X=0.
        \end{equation*}
    \end{itemize}
\end{lemma}

\begin{proof}
    For the no-crossing rule, block diagonality of $H(t)$ implies that
    its unitary propagator $U(t,s)$ commutes with both $P_{<j}$ and
    $P_{\geq j}$. Thus, for any $P,Q\in\{P_{<j},P_{\geq j}\}$,
    \begin{equation*}
        U(t,s)(PXQ)U(t,s)^\dagger
        =P\bigl(U(t,s)XU(t,s)^\dagger\bigr)Q.
    \end{equation*}
    The four choices of $P$ and $Q$ show that each operator block is
    preserved. This also preserves support below the cut and the sum
    of the two off-diagonal blocks.

    For the crossing rule, the transition Hamiltonian itself is
    off-diagonal relative to the cut:
    \begin{equation*}
        H_j(t)
        =
        P_{<j}H_j(t)P_{\geq j}
        +
        P_{\geq j}H_j(t)P_{<j},
    \end{equation*}
    because this transition only couples the local levels
    $\ket{j-1}$ and $\ket{j}$. Multiplying an off-diagonal operator by a
    diagonal operator, from either side, gives an off-diagonal operator.
    Hence, if $X=X_{\mathrm{diag}(j)}$, both
    $H_j(t)X$ and $XH_j(t)$ are off-diagonal, and so their
    commutator $\calL_j(t)[X]$ is off-diagonal as well.

    For trace cancellation, let $X=X_{\mathrm{off}(j)}$ be trace-class.
    Since the projections act only on the bath, cyclicity of the bath
    partial trace gives
    \begin{equation*}
        \tr_b(P_{<j}XP_{\geq j})
        =\tr_b(P_{\geq j}P_{<j}X)=0.
    \end{equation*}
    The other off-diagonal block has zero bath partial trace by the
    same argument. Adding the two contributions gives $\tr_b X=0$.
\end{proof}

We first prove the first-crossing formula by descending induction on the
transition label. With $\calL_j=\calL_{n,r,j}$ and
$\rho_0=\rho_{sb}(0)$, the identity to be proved in
\cref{lem:first-crossing-expansion} is
\begin{align*}
    \calE_{n,r}^{(d)}(t)[\rho_s(0)]
    &=\int_{0\leq t_1\leq\cdots\leq t_d\leq t}
    \tr_b\Bigl[
        \calV_d(t,t_d)\calL_d(t_d)\cdots
        \calV_j(t_{j+1},t_j)\calL_j(t_j)\cdots \\
    &\hspace{25mm}
        \calV_1(t_2,t_1)\calL_1(t_1) \,
        \calV_0(t_1,0)[\rho_0]
    \Bigr]\,\rd t_1\cdots\rd t_d .
\end{align*}
The factors alternate between propagators $\calV_j$ and transitions
$\calL_j$, with the transition labels increasing from right to left.

\begin{proof}[Proof of \cref{lem:first-crossing-expansion}]
    Using $\calV_d=\calU_{\dil}$ and $\calV_{d-1}=\calU_{\dil,d}$, the
    boundary-reduction formula in \cref{lem:diluted-boundary-reduction}
    becomes
    \begin{equation*}
        \calE_{n,r}^{(d)}(t)[\rho_s(0)]
        =\int_0^t
        \tr_b\Bigl[
            \calV_d(t,t_d)\calL_d(t_d)
            \calV_{d-1}(t_d,0)[\rho_0]
        \Bigr]\,\rd t_d .
    \end{equation*}
    This exposes transition $d$ and starts the induction at $j=d-1$.
    Next, we repeatedly apply Duhamel's identity to the rightmost
    propagator $\calV_{j}$, exposing one transition $\calL_j$ at each
    step for $j=d-1,\ldots,1$. At stage $j$, suppose that transitions
    $j+1,\ldots,d$ have been exposed in chronological order, with
    rightmost factor $\calV_j(t_{j+1},0)[\rho_0]$ immediately to the
    right of $\calL_{j+1}(t_{j+1})$.

    Since the generators of $\calV_j$ and $\calV_{j-1}$ differ by
    $\calL_j$, Duhamel's identity gives
    \begin{align*}
        \calV_j(t_{j+1},0)
        =\calV_{j-1}(t_{j+1},0)
        +\int_0^{t_{j+1}}
        \calV_j(t_{j+1},s)\calL_j(s)
        \calV_{j-1}(s,0)\,\rd s .
    \end{align*}
    The first term gives zero after acting on $\rho_0$ and applying
    $\calL_{j+1}$. Indeed, $\calV_{j-1}$ excludes $H_j$, so the
    no-crossing rule in \cref{lem:cut-invariance} keeps the evolved
    operator supported below the cut at $j-1\leftrightarrow j$, since
    the copied mode is initially in vacuum. The Hamiltonian $H_{j+1}$
    acts entirely above that cut, so its commutator with this operator
    vanishes. Thus
    \begin{equation*}
        \calL_{j+1}(t_{j+1})
        \calV_{j-1}(t_{j+1},0)[\rho_0]=0.
    \end{equation*}
    In the surviving integral, rename $s$ as $t_j$. This exposes
    $\calL_j(t_j)$ and leaves $\calV_{j-1}(t_j,0)[\rho_0]$ as the
    rightmost factor, establishing the next induction stage. Iterating
    down to $j=1$ gives the claimed expansion on $0\leq
    t_1\leq\cdots\leq t_d\leq t$.
\end{proof}

\begin{proof}[Proof of \cref{lem:second-crossing-expansion}]
    Start from the first-crossing formula in
    \cref{lem:first-crossing-expansion}:
    \begin{align*}
        \calE_{n,r}^{(d)}(t)[\rho_s(0)]
        &=\int_{0\leq t_1\leq\cdots\leq t_d\leq t}
        \tr_b\Bigl[
            \calV_d(t,t_d)\calL_d(t_d) \times 
            \calV_{d-1}(t_d,t_{d-1})\calL_{d-1}(t_{d-1}) \times \cdots \\
        &\qquad\qquad\qquad\qquad\cdots
            \calV_1(t_2,t_1)\calL_1(t_1) \times
            \calV_0(t_1,0)[\rho_0]
        \Bigr]\,\rd t_1\cdots\rd t_d .
    \end{align*}
    The corresponding transition word for the Liouvillians is
    $(1,\ldots,d)$. Now we add a second occurrence of the Liouvillian
    $\calL_j$ in descending order $j=d,d-1,\ldots,1$, each gives a
    dilution factor $q^{-1/2}$. 

    \paragraph{Starting step.}
    For $j=d$, the only propagator left to $\calL_d$ is
    $\calV_d(t,t_d)$. Removing $H_d$ and applying Duhamel's identity
    gives
    \begin{equation} \label{eq:second-d-expansion}
        \calV_d(t,t_d)
        =\calV_d^{(-\{d\})}(t,t_d)
        +\int_{t_d}^{t}
        \calV_d(t,\tau)\calL_d(\tau)
        \calV_d^{(-\{d\})}(\tau,t_d)\,\rd\tau,
    \end{equation}
    where the propagator $\calV_d^{(-\{d\})}$ is generated by
    $H_{\dil}-H_d$. Note that the operator right to $\calL_d$
    \begin{equation*}
        \calV_{d-1}(t_d,t_{d-1})\calL_{d-1}(t_{d-1}) \times \cdots \times
        \calV_1(t_2,t_1)\calL_1(t_1) \times
        \calV_0(t_1,0) [\rho_0]
    \end{equation*}
    is supported below the $d$-cut with the initial vacuum: the earlier
    propagators exclude $H_d$, and the earlier explicit transitions have
    labels less than $d$. Applying $\calL_d(t_d)$ makes it off-diagonal
    across that cut. Thus, the first term in
    \eqref{eq:second-d-expansion} preserves this off-diagonal sector, so
    its contribution vanishes under the bath trace by
    \cref{lem:cut-invariance}. The integral term in
    \eqref{eq:second-d-expansion} then gives
    \begin{align*}
        \calE_{n,r}^{(d)}(t)[\rho_s(0)]
        &=\int_{0\leq t_1\leq\cdots\leq t_d\leq\tau\leq t}
        \tr_b\Bigl[
            \calV_d(t,\tau)\calL_d(\tau)
            \calV_d^{(-\{d\})}(\tau,t_d) \times \calL_d(t_d)\\
        &\times
            \calV_{d-1}(t_d,t_{d-1})\calL_{d-1}(t_{d-1})\cdots
            \calV_1(t_2,t_1)\calL_1(t_1)
        \times
            \calV_0(t_1,0)[\rho_0]
        \Bigr]\,\rd t_1\cdots\rd t_d\,\rd\tau .
    \end{align*}
    The new transition word for Liouvillians is $(1,\ldots,d-1,d,d)$. 

    \paragraph{Next step.}
    We now expose a second $\calL_{d-1}$ insertion. After the starting
    step, the operator just before the first $\calL_{d-1}$ occurrence is
    \begin{equation*}
        Y:=
        \calV_{d-2}(t_{d-1},t_{d-2})\calL_{d-2}(t_{d-2}) 
        \times \cdots \times
        \calV_1(t_2,t_1)\calL_1(t_1) \times
        \calV_0(t_1,0) [\rho_0]
    \end{equation*}
    Note that the propagators in $Y$ exclude $H_{d-1}$, and the
    transitions there have labels below $d-1$. Hence the no-crossing
    rule gives $Y=P_{<d-1}YP_{<d-1}$, and $\calL_{d-1}(t_{d-1})[Y]$ is
    off-diagonal across the $(d-1)$-cut.

    Set $t_{d+1}:=\tau$ and $t_{d+2}:=t$. After the starting step, the
    evolution after the first occurrence of $\calL_{d-1}$ contains three
    propagators whose generators contain $H_{d-1}$:
    \begin{align*}
        &\calW_{d+1}\calA_{d+1}\calW_d\calA_d\calW_{d-1}\\
        &\quad:=
        \calV_d(t_{d+2},t_{d+1})\calL_d(t_{d+1}) \times
        \calV_d^{(-\{d\})}(t_{d+1},t_d) \calL_d(t_d) \times
        \calV_{d-1}(t_d,t_{d-1}).
    \end{align*}
    Here $\calA_d=\calL_d(t_d)$ and $\calA_{d+1}=\calL_d(t_{d+1})$,
    while $\calW_k=\calW_k(t_{k+1},t_k)$ for $k=d-1,d,d+1$. Let
    $\calW_k^{(-\{d-1\})}$ be the corresponding propagator with
    $H_{d-1}$ removed, and write
    \begin{equation*}
        \calR_k:=\int_{t_k}^{t_{k+1}}
        \calW_k(t_{k+1},v)\calL_{d-1}(v)
        \calW_k^{(-\{d-1\})}(v,t_k)\,\rd v,
        \qquad k=d-1,d,d+1.
    \end{equation*}
    Since all three generators contain $H_{d-1}$, applying the
    telescoping Duhamel expansion in \cref{lem:telescoping-duhamel}
    gives
    \begin{align*}
        &\calW_{d+1}\calA_{d+1}\calW_d\calA_d\calW_{d-1}\\
        &\quad=
        \calW_{d+1}^{(-\{d-1\})}\calA_{d+1}
        \calW_d^{(-\{d-1\})}\calA_d
        \calW_{d-1}^{(-\{d-1\})}\\
        &\qquad+
        \calW_{d+1}\calA_{d+1}\calW_d\calA_d\calR_{d-1}\\
        &\qquad+
        \calW_{d+1}\calA_{d+1}\calR_d\calA_d
        \calW_{d-1}^{(-\{d-1\})}\\
        &\qquad+
        \calR_{d+1}\calA_{d+1}
        \calW_d^{(-\{d-1\})}\calA_d
        \calW_{d-1}^{(-\{d-1\})}.
    \end{align*}

    The first term gives zero bath partial trace after acting on
    $\calL_{d-1}(t_{d-1})[Y]$. Indeed, each modified propagator
    preserves the off-diagonal sector across the $(d-1)$-cut. The two
    $\calL_d$ insertions also preserve it, because $H_d$ acts entirely
    above that cut. Thus \cref{lem:cut-invariance} gives
    \begin{equation*}
        \tr_b\Bigl[
            \calW_{d+1}^{(-\{d-1\})}\calA_{d+1}
            \calW_d^{(-\{d-1\})}\calA_d
            \calW_{d-1}^{(-\{d-1\})}
            \calL_{d-1}(t_{d-1})[Y]
        \Bigr]=0.
    \end{equation*}
    Each remaining term exposes a second $\calL_{d-1}(v)$. All
    propagators between the first and second occurrences of
    $\calL_{d-1}$ exclude $H_{d-1}$; the later propagators retain their
    original generators. Restoring the existing time integrals gives the
    three chronological transition words for the Liouvillians
    \begin{align*}
        (1, 2, \cdots, d-1, d-1, d, d), \,
        (1, 2, \cdots, d-1, d, d-1, d), \,
        (1, 2, \cdots, d-1, d, d, d-1).
    \end{align*}
    These words represent all three possible placements of a second
    $d-1$ after its first occurrence in the previous transition word
    $(1,2,\ldots,d-1,d,d)$. Labels $d-1$ and $d$ now each occur twice,
    while the lower labels still occur once. For $d=2$, these are
    exactly the three admissible words, completing the expansion. For
    $d>2$, we repeat this step for the lower labels as follows.

    \paragraph{Induction step.}
    Suppose that second occurrences of labels $j+1,\ldots,d$ have
    already been exposed. Fix one resulting term and write its
    transition word of Liouvillians as
    \begin{equation*}
        \omega=(1,\ldots,j,\omega_{j+1},\ldots,\omega_M),
        \qquad M := j+2(d-j) = 2d-j.
    \end{equation*}
    Each higher label $j+1,\ldots,d$ occurs twice, and their first
    occurrences remain ordered. Denote the ordered insertion times by
    $t_1,\ldots,t_M$ and set $t_{M+1}:=t$.

    The operator just before the first $\calL_j$ occurrence is
    \begin{equation*}
        Y_j:=
        \calV_{j-1}(t_j,t_{j-1})\calL_{j-1}(t_{j-1})
        \cdots\calV_1(t_2,t_1)\calL_1(t_1)
        \calV_0(t_1,0)[\rho_0].
    \end{equation*}
    For $j=1$, this is $Y_1=\calV_0(t_1,0)[\rho_0]$. This prefix of the
    first-crossing formula is unchanged by the preceding steps, which
    add second occurrences of higher labels after the first $j$. Its
    propagators exclude $H_j$, and its explicit transitions have labels
    below $j$. Since the copied mode starts in vacuum, the no-crossing
    rule gives
    \begin{equation*}
        Y_j=P_{<j}Y_jP_{<j}.
    \end{equation*}
    Thus $\calL_j(t_j)[Y_j]$ is off-diagonal across the $j$-cut.

    The evolution after the first $\calL_j$ insertion contains
    $M-j+1=2(d-j)+1$ propagators. For $j\leq k\leq M$, let
    $\calW_k(u,v)$ denote evolution from $v$ to $u$ within the interval
    $[t_k,t_{k+1}]$ of the chosen term. Its Hamiltonian is obtained from
    $H_{\dil}$ or $H_{\dil,d}$ by removing some transitions
    $H_{j+1},\ldots,H_d$ of the fixed copied mode, as determined by the
    preceding expansion steps. Thus every generator still contains
    $H_j$. Abbreviate
    \begin{equation*}
        \calW_k:=\calW_k(t_{k+1},t_k),\quad j\leq k\leq M,
        \qquad
        \calA_k:=\calL_{\omega_k}(t_k),\quad j+1\leq k\leq M,
    \end{equation*}
    for the propagators and later explicit insertions. The full
    integrand of the chosen term is then
    \begin{equation*}
        \tr_b\Bigl[
            \calW_M\calA_M\cdots
            \calW_{j+1}\calA_{j+1}\calW_j
            \calL_j(t_j)[Y_j]
        \Bigr].
    \end{equation*}
    Let $\calW_k^{(-\{j\})}$ be the corresponding propagator with $H_j$
    removed, and define
    \begin{equation*}
        \calR_k:=\int_{t_k}^{t_{k+1}}
        \calW_k(t_{k+1},v)\calL_j(v)
        \calW_k^{(-\{j\})}(v,t_k)\,\rd v,
        \qquad j\leq k\leq M.
    \end{equation*}
    Applying the telescoping Duhamel expansion in
    \cref{lem:telescoping-duhamel} to these propagators gives
    \begin{align*}
        &\calW_M\calA_M\cdots
            \calW_{j+1}\calA_{j+1}\calW_j\\
        &\quad=
        \calW_M^{(-\{j\})}\calA_M\cdots
            \calW_{j+1}^{(-\{j\})}\calA_{j+1}
            \calW_j^{(-\{j\})}\\
        &\qquad+\sum_{k=j}^{M}
        \bigl(\calW_M\calA_M\cdots
            \calW_{k+1}\calA_{k+1}\bigr)
        \calR_k
        \bigl(\calA_k\calW_{k-1}^{(-\{j\})}\cdots
            \calA_{j+1}\calW_j^{(-\{j\})}\bigr).
    \end{align*}
    The first parenthesized product is the identity for $k=M$, and the
    last is the identity for $k=j$.

    The first term gives zero bath partial trace after acting on
    $\calL_j(t_j)[Y_j]$. Each modified propagator preserves the
    off-diagonal sector across the $j$-cut. Every later explicit
    transition has label $\omega_k>j$, so its Hamiltonian acts entirely
    above that cut and its Liouvillian also preserves this sector. Hence
    \cref{lem:cut-invariance} gives
    \begin{equation*}
        \tr_b\Bigl[
            \calW_M^{(-\{j\})}\calA_M\cdots
            \calW_{j+1}^{(-\{j\})}\calA_{j+1}
            \calW_j^{(-\{j\})}\calL_j(t_j)[Y_j]
        \Bigr]=0.
    \end{equation*}
    Each remaining term exposes a second $\calL_j(v)$ in the gap
    $t_k\leq v\leq t_{k+1}$. All propagators between the first and
    second insertions exclude $H_j$; the later propagators retain their
    generators. Restoring the existing time integrals and placing $v$ in
    its chosen gap gives the ordered domain for the new word
    \begin{equation*}
        (1,\ldots,j,\omega_{j+1},\ldots,\omega_k,
            j,\omega_{k+1},\ldots,\omega_M),
        \qquad j\leq k\leq M,
    \end{equation*}
    where either part of the higher-label suffix may be empty. These are
    all $2(d-j)+1$ possible placements of a second $j$ after its first
    occurrence. Only $H_j$ is removed from additional generators, so the
    same argument applies to the next label $j-1$.

    \paragraph{Completion.}
    Iterating down to $j=1$ gives two explicit insertions of every
    label, with the first occurrences ordered. Thus every resulting word
    belongs to $\Omega_d$. Conversely, deleting the second occurrences
    in increasing label order uniquely reverses the construction for any
    admissible word. Hence each admissible word is obtained exactly
    once. Including the starting step and the next step, the number of
    terms is
    \begin{equation*}
        \prod_{j=1}^{d}\bigl(2(d-j)+1\bigr)=(2d-1)!!.
    \end{equation*}
    Each term has $2d$ chronologically ordered occurrences of
    Liouvillians and propagators generated by the allowed Hamiltonians
    with local transitions removed. Relabeling its times as
    $t_1,\ldots,t_{2d}$ gives the expansion in
    \cref{lem:second-crossing-expansion}, with at most $(2d-1)!!$
    nonzero terms.
\end{proof}

\section{Trotterization error}
\label{app:trotterization-error}

This appendix proves the error bound for the midpoint--Strang scheme
in \cref{sec:digital-simulation}. Fix $N,q,d$ and retain the notation
$H(t)$, $U(T)$, and $\widehat U(T)$ from that section. The exact and
digital evolutions start from the same state,
\begin{equation*}
    \rho_{N,q,d}(0)
    =
    \rho_s(0)\otimes
    \bigotimes_{n=1}^{N}\bigotimes_{r=1}^{q}\ket{0}\bra{0}_{n,r}.
\end{equation*}
The exact and digitally simulated reduced states are
\begin{equation} \label{eq:finite-bath-reduced-states}
    \begin{aligned}
        \rho_s^{(N,q,d)}(T)
        &:=
        \tr_b\!\left[
            U(T)
            \rho_{N,q,d}(0)
            U(T)^\dagger
        \right],\\
        \widehat\rho_s(T)
        &:=
        \tr_b\!\left[
            \widehat U(T)
            \rho_{N,q,d}(0)
            \widehat U(T)^\dagger
        \right].
    \end{aligned}
\end{equation}
The bath interaction picture leaves these reduced system states
unchanged.

To state the error coefficient, we collect the coupling amplitudes and
their first two time derivatives in the quantities
\begin{equation}\label{eq:digital-coupling-scales}
    \gamma_j:=\opnorm{S}\sup_{t\in[0,T]}
        \left(\sum_{n=1}^N|g_n^{(j)}(t)|^2\right)^{1/2},
    \qquad j=0,1,2.
\end{equation}
Here $g_n^{(0)}=g_n$. Thus $\gamma_0$ measures the collective
interaction strength, while $\gamma_1$ and $\gamma_2$ measure the first
and second time derivatives of the couplings. For notation simplicity,
define the corresponding occupation-growth factor over the interval
$[0,T]$ as
\begin{equation}\label{eq:digital-occupation-factor}
    \eta_T:=1+T\gamma_0.
\end{equation}
Both $\gamma_j$ and $\eta_T$ are independent of the dilution parameter
$q$ and the local cutoff $d$.

We write the digital-error coefficient as the sum of its three sources:
\begin{equation}\label{eq:digital-error-constant}
    R_{\trot}:=R_{\mathrm{mid}}+R_{\mathrm{out}}+R_{\mathrm{in}}.
\end{equation}
The midpoint and outer-Strang contributions are
\begin{equation*}
    \begin{aligned}
    R_{\mathrm{mid}}&:=
        \frac{\gamma_2\eta_T}{6}
        +\frac{4}{3}(\opnorm{H_s}+2\gamma_0)\gamma_1\eta_T^2,
        \\
    R_{\mathrm{out}}&:=
        \frac{2}{3}\opnorm{H_s}^2\gamma_0\eta_T
        +\frac{16}{3}\opnorm{H_s}\gamma_0^2\eta_T^2.
    \end{aligned}
\end{equation*}
The inner Strang contribution controls separating even and odd
transitions within each bath register,
\begin{equation}\label{eq:digital-inner-vacuum}
    R_{\mathrm{in}}:=\begin{cases}
        0,&d=2,\\[2pt]
        \displaystyle\frac{\opnorm{S}^3(d-1)^{3/2}}{\sqrt q}
            \sup_{t\in[0,T]}\sum_{n=1}^N|g_n(t)|^3,&d\geq3.
    \end{cases}
\end{equation}
Thus only the inner contribution depends on dilution, and it decreases
as $q^{-1/2}$. 

\begin{theorem}[Second-order digital simulation error from the vacuum]
    \label{thm:second-order-trotter-error}
    Suppose that $g_1,\ldots,g_N$ are twice continuously differentiable
    on $[0,T]$, $q\geq1$, $d\geq2$, and all bath registers start in the
    vacuum. For any initial system state, the midpoint--Strang circuit
    \eqref{eq:fully-decomposed-step} satisfies
    \begin{equation*}
        \tracenorm{\rho_s^{(N,q,d)}(T)-\widehat\rho_s(T)}
        \leq T(\Delta t)^2R_{\trot},
        \qquad \Delta t=T/M_{\trot}.
    \end{equation*}
    For fixed $N,d,T$ and retained couplings, $R_{\trot}$ is
    nonincreasing in $q$, and independent of $q$ when $d=2$.
\end{theorem}

The bound is formulated using the vacuum initial condition rather than a
full operator-norm estimate on the truncated bath. A direct
operator-norm bound can grow with the number of diluted registers and
would obscure the fact that dilution preserves the collective coupling.
The state-dependent estimate removes this artificial $q$-dependence from
the midpoint and outer-Strang contributions. The remaining $q^{-1/2}$
dependence comes from the inner even--odd splitting within each bath
register.

We first introduce three lemmas used in the proof of
\cref{thm:second-order-trotter-error}. Index the $d$-level bath
registers by $\ell$, with truncated annihilation operators
$a_\ell^{(d)}$. Define the total occupation and bath field with
coefficients $c_{\ell}$ by
\begin{equation} \label{eq:total_occupation_bath_field}
    K:=\sum_\ell(a_\ell^{(d)})^\dagger a_\ell^{(d)},
    \qquad 
    B_c:=\sum_\ell\left(
        c_\ell a_\ell^{(d)}
        +c_\ell^*(a_\ell^{(d)})^\dagger\right).
\end{equation}

\begin{remark}[Identity factors for bath operators]
    Throughout the remainder of this appendix, bath operators acting
    on the joint system--bath space are implicitly tensored with the
    identity $I_s$ on the system. Thus, for any bath operator $A$ and
    joint vector $v$, the expression $Av$ denotes $(I_s\otimes A)v$.
    Scalar additions to operators are understood as multiples of the
    identity on the relevant space. Nontrivial system factors are
    always written explicitly.
\end{remark}

\begin{lemma}[Bath-field bounds and occupation growth]
    \label{lem:digital-bath-bounds}
    For any joint system--bath vector $v$ and total occupation $K$ and
    bath field $B_c$ defined in \eqref{eq:total_occupation_bath_field}
    we have
    \begin{equation}\label{eq:digital-bath-field-bounds}
        \|B_cv\|\leq2\|c\|_2\|(K+1)^{1/2}v\|,
        \qquad
        \|(K+1)^{1/2}B_cv\|\leq2\|c\|_2\|(K+2)v\|.
    \end{equation}
    Suppose a normalized state $\psi(t)$ evolves from a vacuum bath
    under the Hamiltonian
    \begin{equation*}
        H(t):=H_s(t)\otimes I+S(t)\otimes B_{c(t)},
    \end{equation*}
    where $H_s(t)$ and $S(t)$ are Hermitian system operators and all
    coefficients are piecewise continuous. Define
    $\gamma(t):=\|S(t)\|\|c(t)\|_2$. Then
    \begin{equation}\label{eq:digital-vacuum-moments}
        \|(K+1)\psi(t)\|
        \leq\left(1+\int_0^t\gamma(s)\,\rd s\right)^2.
    \end{equation}
\end{lemma}

\begin{proof}
    Write $L_c:=\sum_\ell c_\ell a_\ell^{(d)}$, so that
    $B_c=L_c+L_c^\dagger$. Cauchy--Schwarz gives the annihilation bound
    \begin{equation}\label{eq:digital-annihilation-bound}
        \|L_cv\|^2
        \leq\|c\|_2^2\sum_\ell\|a_\ell^{(d)}v\|^2
        =\|c\|_2^2\langle v,Kv\rangle.
    \end{equation}
    For the creation bound, direct multiplication on one truncated
    register gives
    \begin{equation*}
        [a_\ell^{(d)},(a_\ell^{(d)})^\dagger]
        =I_{d,\ell}-d\ket{d-1}\bra{d-1}_\ell,
    \end{equation*}
    where $I_{d,\ell}=\sum_{j=0}^{d-1}\ket{j}\bra{j}_\ell$. Since each
    register is restricted to
    $\operatorname{span}\{\ket{0},\ldots,\ket{d-1}\}$, we identify local
    operators with their tensor extensions by the identity on all other
    registers, and write $I_b$ for the identity on the joint bath space.
    Hence
    \begin{equation*}
        [L_c,L_c^\dagger]
        =\sum_\ell |c_\ell|^2
            \bigl(I_b-d\ket{d-1}\bra{d-1}_\ell\bigr)
        \leq\|c\|_2^2 I_b,
    \end{equation*}
    since operators on distinct registers commute. The operator
    inequality is understood in the quadratic-form sense.
    Consequently,
    \begin{equation}\label{eq:digital-creation-bound}
        \|L_c^\dagger v\|^2
        =\|L_cv\|^2
            +\langle v,[L_c,L_c^\dagger]v\rangle
        \leq\|c\|_2^2\langle v,(K+1)v\rangle.
    \end{equation}

    The first bound in \eqref{eq:digital-bath-field-bounds} follows
    by the triangle inequality. For the second weighted bound, put
    $n_\ell=(a_\ell^{(d)})^\dagger a_\ell^{(d)}$. On one register,
    $[n_\ell,a_\ell^{(d)}]=-a_\ell^{(d)}$, while operators on
    distinct registers commute. Hence
    \begin{equation*}
        [K,L_c]=-L_c,\qquad [K,L_c^\dagger]=L_c^\dagger.
    \end{equation*}
    Note that ladder $L_c$ lowers total occupation by one, while
    $L_c^\dagger$ raises it by one. We have
    \begin{equation*}
        (K+1)^{1/2}L_c=L_cK^{1/2},
        \qquad
        (K+1)^{1/2}L_c^\dagger=L_c^\dagger(K+2)^{1/2},
    \end{equation*}
    which can be verified on each eigenspace of the occupation operator
    $K$. Taking square roots in \eqref{eq:digital-annihilation-bound}
    and \eqref{eq:digital-creation-bound}, and applying the resulting
    estimates to the weighted input vectors, yields
    \begin{align*}
        \|(K+1)^{1/2}L_cv\|
        &\leq\|c\|_2\|Kv\|,\\
        \|(K+1)^{1/2}L_c^\dagger v\|
        &\leq\|c\|_2\|[(K+1)(K+2)]^{1/2}v\|
        \leq\|c\|_2\|(K+2)v\|.
    \end{align*}
    Since $\|Kv\|\leq\|(K+2)v\|$, the triangle
    inequality proves the second field bound.

    To control occupation growth, set $w(t)=(K+1)\psi(t)$ and
    $y(t)=\|w(t)\|$. Since $K\geq0$ and $\|\psi(t)\|=1$,
    we have $y(t)\geq1$. On each interval of continuity of the
    Hamiltonian, Schr\"odinger's equation gives
    \begin{equation*}
        \dot w=-\ri(K+1)H(t)\psi
        =-\ri H(t)w-\ri[K,H(t)]\psi.
    \end{equation*}
    Differentiating $y^2=\langle w,w\rangle$ and substituting
    for $\dot w$ yields
    \begin{align*}
        2y\dot y
        =\langle\dot w,w\rangle
            +\langle w,\dot w\rangle
        =\ri\langle[K,H(t)]\psi,w\rangle
            -\ri\langle w,[K,H(t)]\psi\rangle
        \leq2\|w\|\,\|[K,H(t)]\psi\|.
    \end{align*}
    Dividing by $2y=2\|w\|>0$ gives
    $\dot y\leq\|[K,H(t)]\psi\|$.

    Since $K$ acts only on the bath, the system-only term in $H(t)$
    commutes with it. The ladder commutators therefore give
    \begin{equation*}
        [K,H(t)]=S(t)\otimes[K,L_{c(t)}+L_{c(t)}^\dagger]
        =S(t)\otimes\bigl(L_{c(t)}^\dagger-L_{c(t)}\bigr).
    \end{equation*}
    The triangle inequality and
    \eqref{eq:digital-annihilation-bound}--\eqref{eq:digital-creation-bound}
    now imply
    \begin{align*}
        \|[K,H(t)]\psi\|
        \leq\|S(t)\|\bigl(
            \|L_{c(t)}^\dagger\psi\|+\|L_{c(t)}\psi\|\bigr)
        \leq2\gamma(t)\|(K+1)^{1/2}\psi\|.
    \end{align*}
    Finally, normalization and Cauchy--Schwarz give
    \begin{equation*}
        \|(K+1)^{1/2}\psi\|^2
        =\langle\psi,(K+1)\psi\rangle
        \leq\|\psi\|\,\|(K+1)\psi\|=y.
    \end{equation*}
    Combining these estimates gives $\dot y\leq2\gamma(t)\sqrt y$,
    and hence
    \begin{equation*}
        \frac{\rd}{\rd t}\sqrt{y(t)}
        =\frac{\dot y(t)}{2\sqrt{y(t)}}\leq\gamma(t).
    \end{equation*}
    Integrating on each interval and using continuity of $y$ at the
    junctions yields
    \begin{equation*}
        \sqrt{y(t)}\leq\sqrt{y(0)}
            +\int_0^t\gamma(s)\,\rd s
        =1+\int_0^t\gamma(s)\,\rd s,
    \end{equation*}
    where $y(0)=1$ because the bath starts in the vacuum.
    Squaring this inequality proves \eqref{eq:digital-vacuum-moments}.
    Continuity of $y$ at segment boundaries allows the same estimate to
    be applied to any finite concatenation of evolution segments.
\end{proof}

\begin{lemma}[Midpoint error with occupation bounds]
    \label{lem:digital-midpoint-error}
    Let $H(t)$ be a twice continuously differentiable Hermitian
    Hamiltonian on $[\tau,\tau+h]$, with midpoint $m=\tau+h/2$.
    Define the exact propagator and its midpoint approximation by
    \begin{equation*}
        U_h:=U(\tau+h,\tau)
        =\calT\exp\left[-\ri\int_\tau^{\tau+h}H(t)\,\rd t\right],
        \qquad
        V_h:=\re^{-\ri hH(m)}.
    \end{equation*}
    For a normalized initial vector $\psi$ at time $\tau$, define the
    trajectory generated by the frozen Hamiltonian $H(m)$ as
    \begin{equation*}
        \phi(t):=\re^{-\ri(t-\tau)H(m)}\psi,
        \qquad \tau\leq t\leq\tau+h.
    \end{equation*}
    Suppose the following two conditions hold. First, the Hamiltonian and its derivatives
    satisfy the following bounds with nonnegative constants $\Lambda_j$:
    for every vector $v$, every $t\in[\tau,\tau+h]$, and $j=0,1,2$,
    \begin{equation}\label{eq:digital-weighted-fields}
        \begin{aligned}
        \|H^{(j)}(t)v\|\leq \Lambda_j\|(K+1)^{1/2}v\|,
        \qquad
        \|(K+1)^{1/2}H^{(j)}(t)v\|\leq \Lambda_j\|(K+2)v\|.
        \end{aligned}
    \end{equation}
    Second, for some $\eta\geq1$, all intermediate
    states along the frozen trajectory obey
    \begin{equation*}
        \|(K+1)^{1/2}\phi(t)\|\leq\eta,
        \qquad\|(K+2)\phi(t)\|\leq2\eta^2
        \quad(\tau\leq t\leq\tau+h).
    \end{equation*}
    Then the error between the exact and frozen final states satisfies
    \begin{equation}\label{eq:digital-midpoint-estimate}
        \|(U_h-V_h)\psi\|
        \leq h^3\left(
            \frac{\Lambda_2\eta}{24}
            +\frac{\Lambda_0\Lambda_1\eta^2}{3}
        \right).
    \end{equation}
\end{lemma}

\begin{proof}
    Duhamel's formula gives
    \begin{equation*}
        (U_h-V_h)\psi=-\ri\int_\tau^{\tau+h}
        U(\tau+h,t)[H(t)-H(m)]\phi(t)\,\rd t.
    \end{equation*}
    Write $H(t)-H(m)=(t-m)H'(m)+Q(t)$. Taylor's theorem with
    integral remainder gives
    \begin{equation*}
        Q(t)=(t-m)^2\int_0^1(1-\theta)
            H''\bigl(m+\theta(t-m)\bigr)\,\rd\theta,
    \end{equation*}
    which holds on both sides of the midpoint $m$. For each
    fixed $t$, apply the remainder to the vector $\phi(t)$, only the
    time argument of $H''$ varies in the integral. The $j=2$ bound in
    \eqref{eq:digital-weighted-fields} and the assumed occupation bound
    give, for every $0\leq\theta\leq1$,
    \begin{equation*}
        \bigl\|H''\bigl(m+\theta(t-m)\bigr)\phi(t)\bigr\|
        \leq\Lambda_2\|(K+1)^{1/2}\phi(t)\|
        \leq\Lambda_2\eta.
    \end{equation*}
    Taking the norm of the integral therefore yields
    \begin{equation*}
        \|Q(t)\phi(t)\|
        \leq(t-m)^2\int_0^1(1-\theta)\Lambda_2\eta\,\rd\theta
        =\frac{\Lambda_2\eta}{2}(t-m)^2.
    \end{equation*}
    For the linear term set $F(t)=U(\tau+h,t)H'(m)\phi(t)$.
    Differentiation gives
    \begin{equation*}
        F'(t)=\ri U(\tau+h,t)
        \bigl(H(t)H'(m)-H'(m)H(m)\bigr)\phi(t).
    \end{equation*}
    To bound the two products, apply the weighted inequalities in
    \eqref{eq:digital-weighted-fields} we have
    \begin{align*}
        \|H(t)H'(m)\phi(t)\|
        &\leq\Lambda_0\|(K+1)^{1/2}H'(m)\phi(t)\|
        \leq\Lambda_0\Lambda_1\|(K+2)\phi(t)\|,\\
        \|H'(m)H(m)\phi(t)\|
        &\leq\Lambda_1\|(K+1)^{1/2}H(m)\phi(t)\|
        \leq\Lambda_1\Lambda_0\|(K+2)\phi(t)\|.
    \end{align*}
    Since $U(\tau+h,t)$ is unitary and
    $\|(K+2)\phi(t)\|\leq2\eta^2$, the triangle inequality gives
    \begin{equation*}
        \|F'(t)\|\leq2\Lambda_0\Lambda_1\|(K+2)\phi(t)\|
        \leq4\Lambda_0\Lambda_1\eta^2.
    \end{equation*}
    Since $F(m)$ is independent of $t$ and
    $\int_\tau^{\tau+h}(t-m)\,\rd t=0$, the linear Taylor term satisfies
    \begin{equation*}
        \int_\tau^{\tau+h}(t-m)F(t)\,\rd t
        =\int_\tau^{\tau+h}(t-m)\bigl(F(t)-F(m)\bigr)\,\rd t.
    \end{equation*}
    The fundamental theorem of calculus and the bound on $F'$ give
    \begin{equation*}
        \|F(t)-F(m)\|
        =\left\|\int_m^t F'(s)\,\rd s\right\|
        \leq4\Lambda_0\Lambda_1\eta^2|t-m|.
    \end{equation*}
    Hence midpoint cancellation supplies an additional factor of
    $|t-m|$, yielding
    \begin{equation*}
        \left\|\int_\tau^{\tau+h}(t-m)F(t)\,\rd t\right\|
        \leq4\Lambda_0\Lambda_1\eta^2
            \int_\tau^{\tau+h}(t-m)^2\,\rd t.
    \end{equation*}
    Combining this estimate with the bound on $Q(t)\phi(t)$ in
    Duhamel's formula, and using unitarity of $U(\tau+h,t)$, gives
    \begin{equation*}
        \|(U_h-V_h)\psi\|
        \leq\left(\frac{\Lambda_2\eta}{2}+4\Lambda_0\Lambda_1\eta^2\right)
            \int_\tau^{\tau+h}(t-m)^2\,\rd t.
    \end{equation*}
    The integral equals $h^3/12$, which proves
    \eqref{eq:digital-midpoint-estimate}.
\end{proof}

\begin{lemma}[Autonomous Strang error, Proposition~10 in \cite{childs2021theory}]
    \label{lem:autonomous-strang-error}
    Let $A$ and $B$ be bounded Hermitian operators.  Then, for every
    $h\geq0$,
    \begin{equation*}
        \opnorm{
            \re^{-\ri h(A+B)}
            -
            \re^{-\ri hA/2}
            \re^{-\ri hB}
            \re^{-\ri hA/2}
        }
        \leq
        h^3
        \left(
            \frac{1}{24}\opnorm{[A,[A,B]]}
            +
            \frac{1}{12}\opnorm{[B,[A,B]]}
        \right).
    \end{equation*}
\end{lemma}

\begin{proof}[Proof of \cref{thm:second-order-trotter-error}]
    First suppose that the initial system state is pure,
    $\rho_s(0)=\ket{\varphi}\bra{\varphi}$. Let
    $\ket{0_b}:=\bigotimes_{n=1}^N\bigotimes_{r=1}^q\ket{0}_{n,r}$
    denote the joint bath vacuum. Denote $h=\Delta t$ and $t_k=kh$, and
    define
    \begin{equation*}
        \psi_0:=\ket{\varphi}\otimes\ket{0_b},
        \qquad \psi_k:=U(t_k,0)\psi_0.
    \end{equation*}
    Thus $\psi_k$ is the exact joint system--bath state vector at the
    beginning of step $k$. Let $m=m_k$ and
    \begin{equation*}
        U_k:=U(t_{k+1},t_k),\quad V_k:=\re^{-\ri hH(m)},\quad
        \widetilde W_k:=\re^{-\ri hH_0/2}
            \re^{-\ri hH_{\mathrm{int}}(m)}\re^{-\ri hH_0/2}.
    \end{equation*}
    We bound the three differences $U_k-V_k$,
    $V_k-\widetilde W_k$, and $\widetilde W_k-W_k$ on this exact
    input state, then sum the errors over time steps.

    \paragraph{Midpoint evaluation: $R_{\mathrm{mid}}$.}
    We apply \cref{lem:digital-midpoint-error} on $[t_k,t_{k+1}]$
    with Hamiltonian $H(t)$ and initial vector $\psi_k$, to compare
    $U_k$ with $V_k$. We verify its Hamiltonian and occupation
    assumptions in turn using \cref{lem:digital-bath-bounds}.

    We first verify the weighted Hamiltonian bounds
    \eqref{eq:digital-weighted-fields}. For $j=0,1,2$, set
    $c_{n,r}^{(j)}(t):=g_n^{(j)}(t)/\sqrt q$. Then
    \begin{equation*}
        \|c^{(j)}(t)\|_2^2
        =\sum_{n=1}^N\sum_{r=1}^q\frac{|g_n^{(j)}(t)|^2}{q}
        =\sum_{n=1}^N|g_n^{(j)}(t)|^2,
        \qquad \opnorm{S}\|c^{(j)}(t)\|_2\leq\gamma_j.
    \end{equation*}
    Factor the interaction derivative as
    \begin{equation*}
        H_{\mathrm{int}}^{(j)}(t)
        =S\otimes B_{c^{(j)}(t)}
        =(S\otimes I_b)(I_s\otimes B_{c^{(j)}(t)}).
    \end{equation*}
    Since $\opnorm{S\otimes I_b}=\opnorm{S}$, the first field bound
    in \eqref{eq:digital-bath-field-bounds} gives, for every joint vector $v$,
    \begin{align*}
        \|H_{\mathrm{int}}^{(j)}(t)v\|
        &\leq\opnorm{S}\|B_{c^{(j)}(t)}v\|\\
        &\leq2\opnorm{S}\|c^{(j)}(t)\|_2\|(K+1)^{1/2}v\|
        \leq2\gamma_j\|(K+1)^{1/2}v\|.
    \end{align*}
    For the weighted estimate, $S\otimes I_b$ commutes with
    $(K+1)^{1/2}$ because they act on different tensor factors. Moving
    the system operator past this weight and applying the second field
    bound in \eqref{eq:digital-bath-field-bounds} therefore yields
    \begin{align*}
        \|(K+1)^{1/2}H_{\mathrm{int}}^{(j)}(t)v\|
        &=\|(S\otimes I_b)(K+1)^{1/2}B_{c^{(j)}(t)}v\|\\
        &\leq\opnorm{S}\|(K+1)^{1/2}B_{c^{(j)}(t)}v\|\\
        &\leq2\opnorm{S}\|c^{(j)}(t)\|_2\|(K+2)v\|
        \leq2\gamma_j\|(K+2)v\|.
    \end{align*}
    Here $H_{\mathrm{int}}^{(0)}=H_{\mathrm{int}}$. Because
    $H_0=H_s\otimes I_b$ is time independent,
    $H^{(j)}(t)=H_{\mathrm{int}}^{(j)}(t)$ for $j=1,2$. For $j=0$,
    we also bound the system term. Since $H_0$ commutes with $K$ and
    $K\geq0$, we have
    \begin{align*}
        \|H_0v\|
        &\leq\opnorm{H_s}\|v\|
        \leq\opnorm{H_s}\|(K+1)^{1/2}v\|,\\
        \|(K+1)^{1/2}H_0v\|
        &\leq\opnorm{H_s}\|(K+1)^{1/2}v\|
        \leq\opnorm{H_s}\|(K+2)v\|.
    \end{align*}
    Adding the system and interaction bounds by the triangle inequality
    verifies \eqref{eq:digital-weighted-fields} with
    \begin{equation*}
        \Lambda_0=\opnorm{H_s}+2\gamma_0,\qquad \Lambda_1=2\gamma_1,\qquad \Lambda_2=2\gamma_2.
    \end{equation*}
    Next, the frozen states required by the
    \cref{lem:digital-midpoint-error} for midpoint error are
    \begin{equation*}
        \phi_{\mathrm{mid}}(s):=\re^{-\ri sH(m)}\psi_k,
        \qquad 0\leq s\leq h.
    \end{equation*}
    Each is reached from the initial vacuum bath by exact evolution to
    $t_k$, followed by frozen evolution for duration $s$. The exact
    coupling strength is
    $\gamma(t)=\opnorm{S}(\sum_n|g_n(t)|^2)^{1/2}\leq\gamma_0$.
    During the frozen evolution, the couplings remain fixed at their
    midpoint values $g_n(m)$, so the coupling strength is
    $\gamma(m)\leq\gamma_0$.
    Applying the occupation-growth bound in
    \cref{lem:digital-bath-bounds} to this complete path gives
    \begin{equation*}
        \|(K+1)\phi_{\mathrm{mid}}(s)\|
        \leq\left(1+\int_0^{t_k}\gamma(t)\,\rd t+s\gamma_0\right)^2
        \leq(1+T\gamma_0)^2=\eta_T^2,
    \end{equation*}
    where $t_k+s\leq t_{k+1}\leq T$. Normalization,
    Cauchy--Schwarz, and the triangle inequality then imply
    \begin{equation}\label{eq:digital-weighted-moments}
        \begin{aligned}
        \|(K+1)^{1/2}\phi_{\mathrm{mid}}(s)\|^2
        &\leq\|(K+1)\phi_{\mathrm{mid}}(s)\|\leq\eta_T^2,\\
        \|(K+2)\phi_{\mathrm{mid}}(s)\|
        &\leq\|(K+1)\phi_{\mathrm{mid}}(s)\|+1
        \leq2\eta_T^2.
        \end{aligned}
    \end{equation}
    Both assumptions of \cref{lem:digital-midpoint-error} are now
    verified. Applying it with $\eta=\eta_T$ gives
    \begin{align*}
        \|(U_k-V_k)\psi_k\|
        &\leq h^3\left[
            \frac{\Lambda_2\eta_T}{24}
            +\frac{\Lambda_0\Lambda_1\eta_T^2}{3}
        \right]\\
        &=h^3\left[
            \frac{\gamma_2\eta_T}{12}
            +\frac{2}{3}(\opnorm{H_s}+2\gamma_0)\gamma_1\eta_T^2
        \right]
        =\frac{h^3}{2}R_{\mathrm{mid}}.
    \end{align*}

    \paragraph{System--interaction splitting: $R_{\mathrm{out}}$.}
    We compare the frozen evolution with its outer Strang approximation,
    \begin{equation*}
        V_k=\re^{-\ri h(H_0+H_{\mathrm{int}}(m))},
        \qquad
        \widetilde W_k=\re^{-\ri hH_0/2}
            \re^{-\ri hH_{\mathrm{int}}(m)}\re^{-\ri hH_0/2},
    \end{equation*}
    and seek a bound on $\|(V_k-\widetilde W_k)\psi_k\|$.
    To use \cref{lem:digital-midpoint-error}, we express this splitting
    error as a midpoint approximation error in the interaction picture
    of $H_0$. On the centered interval $u\in[-h/2,h/2]$, define
    \begin{equation*}
        J(u):=\re^{\ri uH_0}H_{\mathrm{int}}(m)\re^{-\ri uH_0}.
    \end{equation*}
    The standard interaction-picture propagator identity gives
    \begin{equation*}
        V_k=\re^{-\ri hH_0/2}
        \left(\calT\exp\left[-\ri\int_{-h/2}^{h/2}J(u)\,\rd u\right]\right)
        \re^{-\ri hH_0/2}.
    \end{equation*}
    Centering the interval makes the two endpoint system rotations
    equal. Replacing $J(u)$ in the middle propagator by its midpoint
    value $J(0)=H_{\mathrm{int}}(m)$ gives exactly $\widetilde W_k$.

    The input to the middle propagator is
    $\zeta_k:=\re^{-\ri hH_0/2}\psi_k$. Unitarity of the leftmost
    system rotation therefore gives
    \begin{align*}
        \|(V_k-\widetilde W_k)\psi_k\|
        =\left\|\left(
            \calT\exp\left[-\ri\int_{-h/2}^{h/2}J(u)\,\rd u\right]
            -\re^{-\ri hJ(0)}\right)\zeta_k\right\|.
    \end{align*}
    The right-hand side is the midpoint error for Hamiltonian $J(u)$
    on $[-h/2,h/2]$ with initial vector $\zeta_k$. We now verify the
    two assumptions of \cref{lem:digital-midpoint-error} for this
    evolution.

    To verify the Hamiltonian bounds, write
    $J(u)=S_u\otimes A_{N,q,d}(m)$, where
    $S_u:=\re^{\ri uH_s}S\re^{-\ri uH_s}$.
    Unitary conjugation preserves $\opnorm{S}$, and each derivative
    adds a commutator with $H_s$. Hence
    $\opnorm{S_u^{(j)}}\leq(2\opnorm{H_s})^j\opnorm{S}$ for
    $j=0,1,2$. The bath field is independent of $u$, so
    \begin{equation*}
        J^{(j)}(u)=S_u^{(j)}\otimes B_{c(m)},
        \qquad c_{n,r}(m)=g_n(m)/\sqrt q.
    \end{equation*}
    Applying the field bounds in \eqref{eq:digital-bath-field-bounds} to
    the bath operator $B_{c(m)}$ therefore gives, for every joint vector
    $v$,
    \begin{align*}
        \|J^{(j)}(u)v\|
        &\leq2\opnorm{S_u^{(j)}}\|c(m)\|_2
            \|(K+1)^{1/2}v\|,\\
        \|(K+1)^{1/2}J^{(j)}(u)v\|
        &\leq2\opnorm{S_u^{(j)}}\|c(m)\|_2\|(K+2)v\|.
    \end{align*}
    Since $\opnorm{S}\|c(m)\|_2\leq\gamma_0$, the common coefficient
    is at most $2(2\opnorm{H_s})^j\gamma_0$. Thus
    \eqref{eq:digital-weighted-fields} holds with
    \begin{equation*}
        \Lambda_0=2\gamma_0,\qquad
        \Lambda_1=4\opnorm{H_s}\gamma_0,\qquad
        \Lambda_2=8\opnorm{H_s}^2\gamma_0.
    \end{equation*}

    For the occupation assumption, the required frozen states are
    \begin{equation*}
        \phi_{\mathrm{out}}(s):=\re^{-\ri sJ(0)}\zeta_k
        =\re^{-\ri sH_{\mathrm{int}}(m)}
            \re^{-\ri hH_0/2}\psi_k,
        \qquad 0\leq s\leq h,
    \end{equation*}
    where $s=u+h/2$ is the elapsed time. Their path from the initial
    vacuum consists of exact evolution to $\psi_k$, a system-only
    rotation, and frozen interaction for duration $s$. The rotation
    commutes with $K$ and contributes zero bath coupling; the frozen
    interaction has strength at most $\gamma_0$. The occupation-growth
    bound in \cref{lem:digital-bath-bounds} therefore gives
    \begin{equation*}
        \|(K+1)\phi_{\mathrm{out}}(s)\|
        \leq\left(1+\int_0^{t_k}\gamma(t)\,\rd t+s\gamma_0\right)^2
        \leq\eta_T^2.
    \end{equation*}
    Normalization, Cauchy--Schwarz, and the triangle inequality yield
    \begin{equation*}
        \|(K+1)^{1/2}\phi_{\mathrm{out}}(s)\|\leq\eta_T,
        \qquad
        \|(K+2)\phi_{\mathrm{out}}(s)\|
        \leq\eta_T^2+1\leq2\eta_T^2.
    \end{equation*}
    With both assumptions verified, \cref{lem:digital-midpoint-error}
    applies with $\eta=\eta_T$ and gives
    \begin{align*}
        \|(V_k-\widetilde W_k)\psi_k\|
        &\leq h^3\left[
            \frac{\Lambda_2\eta_T}{24}
            +\frac{\Lambda_0\Lambda_1\eta_T^2}{3}
        \right]\\
        &=h^3\left[
            \frac{\opnorm{H_s}^2\gamma_0\eta_T}{3}
            +\frac{8}{3}\opnorm{H_s}\gamma_0^2\eta_T^2
        \right]
        =\frac{h^3}{2}R_{\mathrm{out}}.
    \end{align*}

    \paragraph{Even--odd splitting: $R_{\mathrm{in}}$.}
    Apply \cref{lem:autonomous-strang-error} to each mode with
    \begin{equation*}
        A=H_{n,r,\even}(m),\qquad B=H_{n,r,\odd}(m).
    \end{equation*}
    This choice matches the even--odd--even ordering in
    \eqref{eq:fully-decomposed-step}.
    This bounds each mode's error by $h^3$ times its two
    nested-commutator norms, with coefficients $1/24$ and $1/12$. Using
    \begin{equation*}
        \opnorm{H_{n,r,p}(m)}\leq
            \opnorm{S}\sqrt{(d-1)/q}\,|g_n(m)|,
        \qquad
        \opnorm{[A,[B,C]]}\leq4\opnorm{A}\opnorm{B}\opnorm{C},
    \end{equation*}
    and summing over the commuting distinct modes gives
    $\opnorm{\widetilde W_k-W_k}\leq h^3R_{\mathrm{in}}/2$.
    For $d=2$ this error vanishes because the even block is absent.

    \paragraph{Accumulation over time steps.}
    Combining the three estimates gives
    $\|(U_k-W_k)\psi_k\|\leq h^3R_{\trot}/2$. Apply the telescoping
    identity in \cref{lem:telescoping} with $X_k=W_k$ and $Y_k=U_k$, and
    then act on $\psi_0$. Since
    $\psi_k=U_{k-1}\cdots U_0\psi_0$, this gives
    \begin{equation*}
        (U(T)-\widehat U(T))\psi_0
        =\sum_{k=0}^{M_{\trot}-1}
          W_{M_{\trot}-1}\cdots W_{k+1}(U_k-W_k)\psi_k.
    \end{equation*}
    Unitarity, the pure-state trace-distance bound, and partial-trace
    contractivity then give
    \begin{equation*}
        \tracenorm{\rho_s^{(N,q,d)}(T)-\widehat\rho_s(T)}
        \leq2\sum_k\|(U_k-W_k)\psi_k\|
        \leq M_{\trot}h^3R_{\trot}=Th^2R_{\trot}.
    \end{equation*}
    Finally, for a mixed initial system state, write
    $\rho_s(0)=\sum_j p_j\ket{\varphi_j}\bra{\varphi_j}$,
    where $p_j\geq0$, $\sum_jp_j=1$, and each $\varphi_j$ is normalized.
    Denote the exact and digital reduced states obtained from
    $\ket{\varphi_j}\bra{\varphi_j}$ and the same vacuum bath by
    $\rho_{s,j}^{(N,q,d)}(T)$ and $\widehat\rho_{s,j}(T)$, respectively.
    Since the pure-state bound is independent of $\varphi_j$, linearity
    of both evolutions and convexity of the trace norm give
    \begin{equation*}
        \tracenorm{\rho_s^{(N,q,d)}(T)-\widehat\rho_s(T)}
        \leq\sum_j p_j
            \tracenorm{\rho_{s,j}^{(N,q,d)}(T)-\widehat\rho_{s,j}(T)}
        \leq Th^2R_{\trot}.
    \end{equation*}
\end{proof}

\section{Detailed Resource Estimates}
\label{app:resource-estimates}

This appendix derives the sufficient parameter scalings used in
\cref{subsec:circuit-resources}. The system operators are fixed,
$T\geq1$ in fixed time units, and $0<\veps<1$. Each error budget is a
fixed positive fraction of $\veps$. All implicit constants are
independent of $T$ and $\veps$.

\subsection{Assumptions and retained rank}

As in the main text, assume that the kernel eigenvalues on $[0,T]$
satisfy
\begin{equation*}
    \lambda_{n,T}\leq C_\lambda T^\kappa n^{-p},
    \qquad p>1,\quad \kappa\geq0,
\end{equation*}
and that the effective couplings are twice continuously differentiable
with
\begin{equation*}
    \Gamma(T):=\max_{0\leq j\leq2}\sup_{n\geq1}
    \sup_{t\in[0,T]}|g_n^{(j)}(t)|
    \leq C_g T^c,\qquad c\geq0.
\end{equation*}
The spectral assumption controls both the retained rank and occupation
parameters, while the coupling envelope controls the digital evolution.

For $N\geq1$, the discarded spectral tail satisfies
\begin{equation*}
    \eta_N
    =T\left(\sum_{n>N}\lambda_{n,T}^2\right)^{1/2}
    \leq C_\lambda T^{\kappa+1}
        \left(\int_N^\infty x^{-2p}\,\rd x\right)^{1/2}
    =\frac{C_\lambda T^{\kappa+1}}{\sqrt{2p-1}}
        N^{-(p-1/2)}.
\end{equation*}
Since $\ln(1+\veps_{\rank})=\Theta(\veps)$, the rank criterion
\eqref{eq:rank-truncation-choice} can be met with
\begin{equation*}
    N=O(T^a\veps^{-b}),\qquad
    a:=\frac{\kappa+1}{p-1/2},\quad b:=\frac{1}{p-1/2}.
\end{equation*}

\subsection{Occupation parameters}

Since $p>1$, the total spectral weight obeys
\begin{equation*}
    \sum_{n\geq1}\lambda_{n,T}
    \leq C_\lambda\zeta(p)T^\kappa,
    \qquad \zeta(p):=\sum_{n\geq1}n^{-p}<\infty.
\end{equation*}
Using $g_n(t)=\sqrt{\lambda_{n,T}}u_n(t)$ and the normalization of
$u_n$, Cauchy--Schwarz inequality gives
\begin{equation*}
    \beta_{n,T}^2
    =\opnorm{S}^2\lambda_{n,T}
        \left(\int_0^T|u_n(t)|\,\rd t\right)^2
    \leq\opnorm{S}^2T\lambda_{n,T}.
\end{equation*}
In particular, $\beta_{\max}^2=O(T^{\kappa+1})$. Substitution into the
direct-cutoff prescription \eqref{eq:resource-direct-choice} gives
\begin{equation*}
    q=1,\qquad
    d=O\bigl(T^{\kappa+1}+\log_2(N/\veps)\bigr)
     =O\bigl(T^{\kappa+1}+\log_2(1/\veps)\bigr).
\end{equation*}
The last equality uses the rank bound and $T\geq1$.

For dilution at a fixed integer $d\geq2$, positivity of the eigenvalues
instead gives
\begin{equation*}
    \begin{aligned}
    \sum_{n=1}^N\beta_{n,T}^{2d}
    \leq\opnorm{S}^{2d}T^d\sum_{n=1}^N\lambda_{n,T}^d
    \leq\opnorm{S}^{2d}T^d
        \left(\sum_{n=1}^N\lambda_{n,T}\right)^d
    \leq\opnorm{S}^{2d}[C_\lambda\zeta(p)]^dT^{d(\kappa+1)}.
    \end{aligned}
\end{equation*}
The dilution prescription \eqref{eq:resource-diluted-choice} therefore
allows
\begin{equation*}
    q=O\left(T^{(\kappa+1)d/(d-1)}\veps^{-1/(d-1)}\right),
    \qquad d\ \text{fixed}.
\end{equation*}

\subsection{Digital evolution and circuit cost}

Define a common scale independent of dilution,
\begin{equation*}
    \Lambda:=1+\sqrt{dN}\,T^c.
\end{equation*}
For fixed system operators, the coupling envelope
$\Gamma(T)\leq C_gT^c$ gives
\begin{equation*}
    \gamma_j\leq\opnorm{S}\sqrt N\,\Gamma(T)=O(\Lambda),
    \quad j=0,1,2,\qquad \eta_T=1+T\gamma_0=O(T\Lambda).
\end{equation*}
The midpoint and outer splitting contributions therefore satisfy
\begin{equation*}
    R_{\mathrm{mid}},\ R_{\mathrm{out}}=O(T^2\Lambda^4).
\end{equation*}
The inner splitting contribution satisfies
\begin{equation*}
    R_{\mathrm{in}}
    =O\left(\frac{(d-1)^{3/2}N\,T^{3c}}{\sqrt q}\right)
    =O(\Lambda^3/\sqrt q),
\end{equation*}
and is zero when $d=2$. Adding the three contributions gives
\begin{equation*}
    R_{\trot}
    =O\left(T^2\Lambda^4+\frac{\Lambda^3}{\sqrt q}\right)
    =O(T^2\Lambda^4),
\end{equation*}
where the final bound is uniform in $q\geq1$.
The step prescription \eqref{eq:resource-step-choice} allows
\begin{equation*}
    M_{\trot}=O\left(T^{5/2}\veps^{-1/2}\Lambda^2\right).
\end{equation*}
The extra time dependence comes from controlling occupation moments
along the evolution. 

Substitution into \cref{eq:resource-width,eq:resource-depth} gives
\begin{equation*}
    \begin{aligned}
    W_{\mathrm{circ}}
    &=n_s+Nq\,w_{\mathsf e}(d),\\
    D_{\mathrm{circ}}
    &=O\left[
        \left(1+T^{5/2}\veps^{-1/2}\Lambda^2\right)
        \left(D_{\mathrm{sys}}+NqdD_{\mathrm{int}}(d,\mathsf e)\right)
    \right].
    \end{aligned}
\end{equation*}
For either binary or unary encoding, the parameter choices above make
$W_{\mathrm{circ}}$ polynomially bounded in $T$ and $1/\veps$. They also
make $\Lambda$ and the step count polynomially bounded, so the circuit
depth is polynomial provided the elementary implementation costs are
polynomially bounded. 

\bibliography{quanEvo.bib}

\end{document}